\documentclass[doublecol]{epl2}

\usepackage{amsthm}
\usepackage{amsmath}

\usepackage{amstext}

\usepackage{dsfont}
\usepackage{bbm}
\usepackage{graphicx}
\usepackage{epstopdf}
\usepackage{subfigure}
\usepackage{dcolumn}
\usepackage{braket}

\usepackage{cite}

\theoremstyle{plain}\newtheorem{theorem}{Theorem}
\theoremstyle{plain}

\title{Complementarity between Success Probability and Coherence in Grover Search Algorithm}

\author{Minghua Pan\inst{1,2,\thanks{Email:\email{panmh@guet.edu.cn}}} \and Haozhen Situ\inst{3,\thanks{Email:\email{situhaozhen@gmail.com}}} \and Shenggen Zheng\inst{4,\thanks{Email:\email{zhengshg@pcl.ac.cn}}}}
\shortauthor{M. Pan,  H. Situ, S. Zheng}

\institute{
  \inst{1} Guangxi Key Laboratory of Cryptography and Information Security, Guilin University of Electronic Technology, Guilin 541004, China\\
  \inst{2} Department of Physics, Tsinghua University, Beijing 100084, China\\
  \inst{3}College of Mathematics and Informatics, South China Agricultural University, Guangzhou 510642, China\\
  \inst{4}Peng Cheng Laboratory, Shenzhen 518055, China
}

\abstract{Coherence plays a very important role in Grover search algorithm (GSA).
In this paper, we define the normalization coherence $\mathds{N}(C)$, where $C$ is a coherence measurement.
In virtue of the constraint of large $N$ and Shannon's maximum entropy principle, a surprising  complementary relationship between the coherence and the success probability of GSA is obtained. Namely,
$P_s(t)+\mathds{N}(C(t))\simeq 1$,
where $C$ is in terms of the relative entropy of coherence and $l_1$ norm of coherence, $t$ is the number of the search iterations in GSA.
Moreover, the equation holds no matter in ideal or noisy environments.
Considering the number of qubits is limited in the recent noisy intermediate-scale quantum (NISQ) era, some exact numerical calculation experiments are presented for different database sizes $N$ with different types of noises. The results show that the complementary between the success probability and the coherence almost always hold.
This work provides a new perspective  to improve the success probability by manipulating its complementary coherence, and vice versa.
It has an excellent potential for helping  quantum algorithms design in the NISQ era.}

\begin{document}

\maketitle

\section{Introduction}
\quad  Quantum computing devices have achieved their superiority in some special problems \cite{Arute,Zhong} since 2019.
However, it is far beyond a general quantum computer which is fault-tolerant and requires millions of qubits with low error rates and long coherence time.
Nowadays, some quantum computing devices which consist of dozens or hundreds of noisy qubits have been obtained. These devices perform imperfect operations within a limited coherent time without error correction and are named as noisy intermediate-scale quantum (NISQ) computers \cite{Preskill}.
It has been shown that quantum algorithms operating on these NISQ devices have advantages in diverse disciplines \cite{Bharti,Havlicek,Huang,McArdle,Situ,ZhangSH22,LiuJ22}.

Grover search algorithm (GSA) \cite{Grover97,Nielsen} is one of the most important quantum algorithms, because of its quadratic acceleration compared to its classical one and has broad applications in a variety of scenarios.
Therefore, GSA has attracted the interest of many researchers and was generalized and improved in various scenarios \cite{Biham02,Long99,Long01,Yoder14,Zhang20}.
It has been shown quantum correlation resources such as entanglement and coherence play important roles in quantum algorithms.
Hence, there are many fruitful works on the quantum correlation resources in GSA \cite{Braunstein,Bru,Chin,Cui,Jozsa,Rossi,Pan17,Rungta,Shi,Pan191,Pan192}.

Although coherence is crucial in quantum algorithms, it was first proposed as a quantifiable quantum resource by Baumgratz et al. \cite{Baumgratz} in 2014. Based on their work, researches on coherence emerged in large numbers \cite{Xi2015,Yuan15,Streltsov15}.
Shi et al.\cite{Shi} found in 2017 that the coherence was depleted in GSA as the success probability increases to the maximum.
In 2019, we studied how the four operators affect the success probability and the coherence in GSA and obtained similar results that the improvement of the success probability was at the expense of coherence consumption \cite{Pan192}.

Noise always affects the efficiency and performance of quantum algorithms. Soon after GSA was proposed, scientists began to study the performance of GSA under different noisy environments \cite{Azuma,Botsinis,Cohn,Gawron,Long,Pablo,Pan21,Rastegin,Reitzner,Salas,Shapira,Shenvi,Wang,Zhirov}.
In \cite{Rastegin,Rastegin21}, Rastegin et al. showed  the trade-off between the success probability and coherence in GSA with collective phase flip and amplitude damping noises on Oracles.

Although the previous works on coherence and noise in GSA have achieved fruitful results, there still lacks a clear and unified characterization between the success probability and the coherence, especially in different noise environments.
To deal with  this problem, we focus on the relationship between the coherence and the success probability in different environments in Bloch representation inspired by Refs.\cite{Rastegin,Pan21}.
In virtue of the constraint of large $N$ and Shannon's maximum entropy principle \cite{Shannon}, the complementary between the success probability $P_s(t)$ of GSA and the normalization of coherence $\mathds{N}(C(t))$ is obtained, where $C$ is in terms of the relative entropy of coherence
 and the $l_1$ norm of coherence
and $t$ is the number of the search iterations.
In a simple and elegant way, we show that $P_s(t)+\mathds{N}(C(t))\simeq 1$.
At the same time, exact numerical experiments are presented for different database sizes with different types of noises. The numerical results show that the above complementary between the success probability and the coherence almost always holds.


\section{Preliminary}\label{Sec2}
In this section, we review Grover's search algorithm (GSA) and some typical noise models and coherence measurement.

\subsection{Grover's search algorithm}\label{GSA}
In 1996, Grover \cite{Grover97} introduced a famous quantum algorithm which named as Grover search algorithm (GSA).
Suppose there is an $N=2^n$ items unstructured database and $M$ targets that satisfy some special conditions, the aim is to search through the database and find out the index of one of the targets.
The process of GSA is as follows (for more details see Refs. \cite{Grover97,Nielsen}).

First, the state of the system is initialized to an equal superposition state
$\ket{\psi_0}=H^{\otimes n}\ket{0}=\frac{1}{\sqrt N}\sum_{x=0}^{N-1}\ket{x}$, where $H$ is a Hadamard matrix.
For convenience, we denote $\ket{\chi_0}=\frac{1}{\sqrt{N-M}}\sum_{x_{n}}\ket{x_{n}}$
and
$\ket{\chi_1}=\frac{1}{\sqrt{M}}\sum_{x_s}\ket{x_s}$ as the superpositions of all the not-target states and all the target states, respectively.
Let $\theta/2=\arcsin\sqrt{M/N}$, the initial state $\ket{\psi_0}$ can be written as
\begin{eqnarray}
\ket{\psi_0} 
=\cos\frac{\theta}{2}\ket{\chi_0}+\sin\frac{\theta}{2}\ket{\chi_1}.
\end{eqnarray}

Then, GSA applies a quantum subroutine \emph{Grover iteration} (\emph{G}), repeatedly.
The subroutine of Grover iteration can be decomposed into four quantum operators $G=H^{\otimes n}S_pH^{\otimes n}O$, where
\begin{itemize}
  \item $O$ is an oracle. It keeps the not-target states unchanged and invert the target states, $$O\ket{x}=(-1)^{f(x)}\ket{x},$$
      where the function $f(x)=1$ if $x$ is an index of target states,
      else $f(x)=0$.

  \item $S_p$ is a conditional phase shift operator. It makes every computational basis state except $\ket{0}$ receiving a phase shift of $-1$, $$S_p\ket{x}=-(-1)^{\delta_{x0}}\ket{x}.$$

\end{itemize}

In the two-dimension space spanning by  $\{\ket{\chi_0},\ket{\chi_1}\}$, the oracle $O$ performs a reflection about $\ket{\chi_0}$ as $O=2\ket{\chi_0}\bra{\chi_0}-I$, where
$I$ is an identity matrix.
Let $R=H^{\otimes n}S_pH^{\otimes n}$.
It was proven that $R=2\ket{\psi_0}\bra{\psi_0}-I$ which is a reflection operator about $\ket{\psi_0}$.
Therefore, we obtain
\begin{eqnarray}\label{psi1}
\ket{\psi_1} = G\ket{\psi_0}=\cos{\frac{3\theta}{2}}\ket{\chi_0}+\sin{\frac{3\theta}{2}}\ket{\chi_1}.
\end{eqnarray}
It suggests that the two reflections produce a rotation with angle $\theta$.

After $t$ repetitions of Grover iteration, the state becomes
\begin{eqnarray}
\ket{\psi_t}= G^t\ket{\psi_0}=\cos\theta_t\ket{\chi_0}+\sin\theta_t\ket{\chi_1},
\end{eqnarray}
with $\theta_t=(2t+1)\frac{\theta}{2}$.
Therefore, the success probability of GSA is $P_s(t)=\sin^2\theta_t$.
The optimal number of iterations is $T=\lfloor\frac{\pi}{4}\sqrt{N/M}\rfloor$ in the case of $M\ll N$.
Hence, it is a quadratic acceleration for GSA compared with its classical algorithms which require $\Omega(N)$ iterations.

\subsection{Noise models}
In the real world, a quantum system inevitably interacts with the environment. That will bring up different types of noises. Quantum operations and the operator-sum representation are used to describe the quantum noise and the behavior of the open quantum systems \cite{Nielsen}. In fact, a \textbf{quantum operation $\varepsilon$} is a map that changes an initial state $\rho$ into the final state $\varepsilon(\rho)$ and can be modeled by operator-sum representation as
\begin{eqnarray}\label{Esum}
\varepsilon(\rho)=\sum_k E_k\rho E_k^\dag,
\end{eqnarray}
where $E_k$ are operation elements and $\sum_k E_k^\dag E_k=I$ if the operation is trace-preserving.
In a two dimensions space, the simplest and most important operations to describe quantum noise models are flip noises, depolarizing noise, and damping noises. Here, we take the flip noises as examples and consider the collective flip noises in GSA.

The \emph{flip noises} include three types of noises, i.e. bit flip, phase flip, and bit-phase flip. For the \emph{bit flip}, like its classic case, it flips a state from $\ket{0}$ to $\ket{1}$ with probability $1-p$ (vice versa from $\ket{1}$ to $\ket{0}$).
For the \emph{phase flip}, it is a quantum-specific noise and brings up a phase flip with $1-p$.
The \emph{bit-phase flip} will cause the state to occur both bit flip and phase flip.
The operation elements are $E_0=\sqrt p I, E_1=\sqrt {1-p} \sigma_i$ where $i=1,2,3$ and $\sigma_i\in\{X,Y,Z\}$ are Pauli matrices for bit flip, bit-phase flip and phase flip, respectively. In that way, these three type of noises also called $X,Y,Z$ noises.
In Bloch representation, the map is
\begin{equation}\label{fr}
\{r_x,r_y,r_z\}\rightarrow \{a r_x, b r_y, c r_z\},
\end{equation}
where $a,b,c=1$ for $X,Y,Z$ noises, else $a,b,c=1-2p$, respectively.

\subsection{Noisy GSA in Bloch Presentation}\label{NGSA}
Bloch sphere is a useful geometric presentation to visualize the state of a qubit.
For a state $\ket{\varphi}=\cos\frac{\theta}{2}\ket{0}+e^{i\varphi}\sin\frac{\theta}{2}\ket{1}$, its  Bloch vector is $\textbf{r}=(r_x,r_y,r_z)=(\sin\theta\cos\varphi,\sin\theta\sin\varphi,\cos\theta)$.
Let $\rho$ be a density matrix of the state $\ket{\varphi}$,
we have
\begin{align}\label{H}
\rho=\frac{1}{2}\left(
\begin{array}{cc}
1+r_z&r_x-i r_y\\
r_x+i r_y&1-r_z\\
\end{array}
\right).
\end{align}

For the initial state $\ket{\psi_0}$ in GSA, its Bloch vector is $r(0)=(\sin\theta,0,\cos\theta)$. Since $r_y(0)=0$, the discussion about GSA can be restricted in two dimensions of $r_x$ and $r_z$.
Grover iteration in the Bloch presentation by $r_x$ and $r_z$ is modified as \cite{Rastegin,Pan21}
\begin{align}\label{G}
G=\left[
\begin{array}{cc}
\cos2\theta&-\sin2\theta\\
\sin2\theta&\cos2\theta
\end{array}
\right].
\end{align}

Consider the noise matrix is diagonalizable\cite{Pan21}, so that the noisy Grover iterative can be written in the form
$$G_n=G\circ E_n,$$
where $E_n$ is the matrix of noise in Bloch presentation.

Let $\eta=1-2p$, the matrixes $E_n$ for bit flip, phase flip and bit-phase flip in $r_x$ and $r_z$ dimensions in Bloch presentation are
$$E_b=\begin{bmatrix} 1 & 0 \\0 & \eta \end{bmatrix},
E_p=\begin{bmatrix} \eta & 0 \\0 & 1\end{bmatrix},
E_{bp}=\eta \begin{bmatrix} 1 & 0 \\0 &1  \end{bmatrix}.$$

After $t$ iterations, the density matrix becomes
\begin{align}\label{rhot}
\rho(t)=G_n^t \rho(0)G_n^{t \dag}=\frac{1}{2}\left(
\begin{array}{cc}
1+r_z(t)&r_x(t)\\
r_x(t)&1-r_z(t)\\
\end{array}
\right).
\end{align}

Therefore, the success probability $P_s(t)$ is rewritten as
\begin{eqnarray}\label{Ps}
P_s(t)=\frac{1-r_z(t)}{2}.
\end{eqnarray}
Hence the success probability $P_s(t)$ only depends on $r_z(t)$.

\subsection{Coherence measurement}
Baumgratz et al.~\cite{Baumgratz} came up with two coherence measurements, the relative entropy of coherence ($C_e$) and the $l_1$ norm of coherence ($C_l)$.
Based on their works, a number of coherence measures, such as the coherence of formation \cite{Yuan15} and geometric coherence \cite{Streltsov15}, have been proposed.
In this work, we choose $C_e$ and $C_l$ to characterize the coherence of GSA.

For a given density matrix $\rho$, the relative entropy of coherence $C_e$ is
\begin{eqnarray}\label{CE1}
C_e(t) := S(\rho_d(t))-S(\rho(t)),
\end{eqnarray}
where $S(x)= -tr(x \ln x)$ is the von Neumann entropy and $\rho_d$ is the state just take the diagonal elements from $\rho$.

For the $l_1$ norm of coherence $C_l$, it is defined as
\begin{eqnarray}\label{CL1}
C_{l}(t) := \sum_{i,j,i\neq j}|\rho_{ij}(t)|.
\end{eqnarray}
That is the summation of the off-diagonal elements of $\rho$.

\section{Coherence dynamics of GSA in different environments}\label{Sec3}
Now, we will discuss the coherence dynamics of GSA as the number of iterations changes. First, the definition of the normalization of a coherence measurement is put forward.

\subsection{Normalization of a coherence measurement}
Since different coherence measurements can just measure part of coherence, it is difficult to do a comparison among them. The most critical problem is that it is hard to build up the relationship between the success probability $P_s\in[0,1]$ and the coherence as different coherence measurements generally have different scales. To deal with this issue, we bring up the definition of the normalization of a coherence measurement.

\newtheorem{def1}{Definition}
\begin{def1}[\textbf{Normalization of a coherence measurement}]\label{Def1}
Let $C$ and $C_m$ be a coherence measurement and its maximum under some specific conditions. Then $\mathds{N}(C):=C/C_m\in [0,1]$ is the normalization of the coherence measurement $C$.
\end{def1}

According to Definition \ref{Def1}, the normalizations $C_e$ and $C_l$ are defined as follows:
\begin{eqnarray}
&&\mathds{N}(C_e(t)):=C_e(t)/C_{e_m},\label{dNCe}\\
&&\mathds{N}(C_l(t)):=C_l(t)/C_{l_m}.\label{dNCl}
\end{eqnarray}

\subsection{The dynamics coherence of GSA in Bloch}
It has been shown in \cite{Shi} that the coherence of ideal GSA  depleted as the number of search iteration increased. We will discuss typical collective flip noises in Bloch representation as in Ref.\cite{Rastegin,Pan21} in this paper.

According to Ref.\cite{Rastegin}, the von Neumann entropy in Eq.(\ref{CE1}) is
\begin{eqnarray}\label{Sr}
S(\rho(t))=-\nu_+(t)\ln\nu_+(t) -\nu_-(t)\ln\nu_-(t),
\end{eqnarray}
where
\begin{eqnarray}\label{nupm}
\nu_\pm(t)=\frac{1\pm \sqrt{r_x(t)^2+r_z(t)^2}}{2},
\end{eqnarray}
\begin{eqnarray}\label{Sd}
S(\rho_{d}(t))\!=\!P_s(t)\ln\frac{M}{P_s(t)}\!+\!(1-P_s(t))\ln\frac{N-M}{1-P_s(t)}.
\end{eqnarray}
According to Eqs.(\ref{dNCe}) and (\ref{Sr})-(\ref{Sd}), the coherence $C_e(t)$ depends on both $r_x(t)$ and $r_z(t)$.

Now, consider the coherence in terms of the $l_1$ norm of coherence in GSA.
In Ref.\cite{Shi,Pan192}, the $l_1$ norm of coherence in GSA is \begin{eqnarray}\label{GCL1}
C_l(t) = (\sqrt{M}\sin\theta_t +\sqrt{N-M}\cos\theta_t)^2-1.
\end{eqnarray}
Let's rewrite it in Bloch representation. According to Eqs.(\ref{rhot})(\ref{CL1}) and considering the number of targets and non-targets, we obtain that
\begin{eqnarray}\label{Cl2}
C_l(t)
=\frac{1}{2}[N+Nr_z(t)+2\sqrt{(N-M)M}r_x(t)].
\end{eqnarray}
Obviously, the coherence $C_l(t)$ also depends on both $r_x(t)$ and $r_z(t)$.

Nevertheless, we can show that the coherence in terms of $C_e(t)$ and $C_l(t)$ can be determined only by $r_z(t)$ in the constraint of large $N$ and Shannon's maximum entropy principle. Furthermore, we obtain the complementary relationship between $P_s(t)$ and $\mathds{N}(C(t))$ in the following theorem.

\newtheorem{theorem1}{\textbf{Theorem}}
\begin{theorem}\label{Th1}
For Grover search algorithm, the normalization of coherence $\mathds{N}(C(t))$ and the success probability $P_s(t)$ satisfy the complementary relationship for large $N$. That is
\begin{eqnarray}
P_s(t)+\mathds{N}(C(t))\simeq 1,
\end{eqnarray}
where $\simeq$ means the algorithm search in large database satisfying $N>>M$, and $C(t)$ is the relative entropy of coherence or the $l_1$ norm of coherence.
\end{theorem}
\begin{proof}
First, let us consider the relative entropy of coherence.
By expanding Eq.(\ref{Sd}), we have
\begin{eqnarray}
S(\rho_{d}(t))=P_s(t)\ln M+(1-P_s(t))\ln(N-M)\nonumber\\
 \!-\!P_s(t)\ln(P_s(t))\!-\!(1-P_s(t))\ln(1-P_s(t)).
\end{eqnarray}

According to maximum entropy principle \cite{Shannon}, the binary Shannon entropy is always no more that $\ln 2$ in two variables system. Therefore, we have
\begin{eqnarray}\label{ign1}
S(\rho(t))=-\nu_+\ln\nu_+ -\nu_-\ln\nu_- \leq\ln 2
\end{eqnarray}
and
\begin{eqnarray}\label{ign2}
\!-\!P_s(t)\ln(P_s(t))\!-\!(1-P_s(t))\ln(1-P_s(t))\leq \ln 2.
\end{eqnarray}
In the condition $N>>M\geq 1$, the above items in Eqs. (\ref{ign1}) and (\ref{ign2}) can be ignored from $C_e(t)$, furthermore we obtain
\begin{eqnarray}\label{Ce2}
C_e(t)\simeq P_s(t)\ln M+(1-P_s(t))\ln(N-M)
\end{eqnarray}
and its maximum
\begin{eqnarray}\label{Cem2}
C_{e_m}\simeq \ln(N-M).
\end{eqnarray}
Thus, the normalization of the relative entropy of coherence in GSA is
\begin{eqnarray}\label{NCe2}
\mathds{N}(C_e(t))
&&\simeq \frac{P_s(t)\ln M+(1-P_s(t))\ln(N-M)}{\ln(N-M)} \nonumber\\
&&\simeq 1-P_s(t).
\end{eqnarray}

Now, let us consider the $l_1$ norm of coherence.
For $N>>M$ and $r_x(t)\in [0,1]$, the third item in Eq.(\ref{Cl2}) can be ignored that
\begin{eqnarray}\label{Cl3}
C_{l}\simeq \frac{N}{2}[1+r_z(t)].
\end{eqnarray}

Since $r_z(t)\in [0,1]$, it is easy to obtain the maximum of $C_{l}$
\begin{eqnarray}\label{Clm3}
C_{l_m}\simeq \frac{N}{2}[1+\max(r_z(t))]=N.
\end{eqnarray}

Therefore, the normalization of the $l_1$ norm of coherence in GSA is
\begin{eqnarray}\label{NCl2}
\mathds{N}(C_l(t))\simeq \frac{1}{2}[1+r_z(t)].
\end{eqnarray}
Obviously, according to Eqs.(\ref{Ps}) and (\ref{NCl2}), we have
\begin{eqnarray}
P_s(t)+\mathds{N}(C_l(t))\simeq 1.
\end{eqnarray}

\end{proof}

Note that $r_x(t)$ and $r_z(t)$ can be used to describe the iteration process of GSA whether there is noise or not. Therefore, Theorem \ref{Th1} holds in both ideal and different noise environments.

\subsection{Coherence in GSA with different types of noises}\label{SSecNC}
To obtain the dynamics of coherence and the success probability of GSA with different noises exactly, we investigate the amount of them in Bloch presentation.
In order to take the advantage of the iterative equation in Bloch, we consider different types of collective flip noises as discussed in Refs. \cite{Rastegin,Pan21}. Once giving $N$ and $M$, the exact analytical expression of coherence in GSA depends entirely on $r_x(t)$ and $r_z(t)$ according to Eqs.(\ref{Ce2}) and (\ref{Cl2}). Hence, we will just evaluate the expressions of $r_x(t)$ and $r_z(t)$ which are determined by the noise Grover iterations $G_n$.

\textbf{For bit flip.} According to Refs.\cite{Pan21}, we have
\begin{align}\label{Gbt}
G_b^t=\frac{\eta^{t/2}}{B}\left[
\begin{array}{cc}
B\cos\phi t + A_-\sin\phi t & \eta\sin\phi t \sin 2\theta\\
-\sin\phi t \sin 2\theta & B \cos\phi t + A_-\sin\phi t
\end{array}
\right].
\end{align}
where $\phi=\arctan(B/A_+)$ and
\begin{eqnarray}
&&A_\pm :=\frac{1\pm \eta}{2}\cos 2\theta,\label{A}\\
&&B:=\left\{
\begin{aligned}
&\sqrt{\eta-A_+^2}, &&if\ \eta\geq A_+^2\\
&\sqrt{A_+^2-\eta}, &&if\ \eta<A_+^2.  \label{B}
\end{aligned}
\right.
\end{eqnarray}
Accordingly, we obtain
\begin{small}
\begin{eqnarray}
 r_x(t)\!=\!\frac{\eta^{t/2}}{B}[(B\cos\phi t\!+\!A_-\sin\phi t)\sin\theta
 \!+\!\eta\sin 2\theta\sin\phi t\cos\theta], \nonumber\\
r_z(t)\!=\!\frac{\eta^{t/2}}{B}[\!-\!\sin 2\theta\sin\theta\sin\phi t
\!+\!(B\cos\phi t\!-\!A_-\sin\phi t)\cos\theta]. \nonumber\\
\end{eqnarray}
\end{small}

Consequently, the success probability and the coherence in GSA will decay exponentially with the noise since both $r_x(t)$ and $r_x(t)$ decay exponentially as $\eta\in[0,1]$.

\textbf{For phase flip.}
The phase flip is one of the most special and important noises in quantum information processing, which no corresponding classical noise.
The iterative equation of $G_p^t$ with phase flip in Bloch is \cite{Rastegin,Pan21}
\begin{align}
G_{p}^t=\frac{\eta^{t/2}}{B}\left[
\begin{array}{cc}
B\cos\phi t - A_-\sin\phi t & \sin\phi t \sin 2\theta\\
-\eta \sin\phi t \sin 2\theta & B \cos\phi t + A_-\sin\phi t
\end{array}
\right].
\end{align}

Then, we obtain
\begin{small}
\begin{eqnarray}
&&r_x(t)\!=\!\frac{\eta^{t/2}}{B} [(B\cos\phi t\!-\!A_- \sin\phi t)\sin\theta
\!+\!\sin 2\theta\sin\phi t\cos\theta],\nonumber\\
&&r_z(t)\!=\!\frac{\eta^{t/2}}{B} [(B\cos\phi t\!+\!A_-\sin\phi t)\cos\theta\!-\!\eta\sin 2\theta\sin\phi t\sin\theta].\nonumber\\
\end{eqnarray}
\end{small}
Similar to the case of bit flip noise, the success probability and the coherence in GSA will decay exponentially with the phase flip noise.

\textbf{For bit-phase flip.}
The noisy Grover iterator $G_{bp}$ with the bit-phase flip becomes \cite{Pan21}
\begin{align}
G_{bp}=\eta\left[
\begin{array}{cc}
\cos 2\theta & \sin 2\theta\\
-\sin 2\theta & \cos 2\theta
\end{array}
\right]
=\eta G.
\end{align}

Then $r_x(t)$ and $r_z(t)$ becomes
\begin{eqnarray}
r_x(t)=\eta^{t}\sin[(2t+1)\theta],\\
r_z(t)=\eta^{t}\cos[(2t+1)\theta].
\end{eqnarray}
In this way, we  obtain easily that the success probability and the coherence in GSA decay exponentially with the noise since $\eta\in[0,1]$.

\section{Numerical results and Discussion}\label{Sec4}
For more convincing and to visualize the above results, we calculate the exact values of the $P_s(t)$, $C_e(t)$, $C_l(t)$ and $1-\mathds{N}(C_e(t)),1-\mathds{N}(C_l(t))$ with different noises for different $N$, then illustrate them in graphics.

Suppose that we search one target in the database with $N$ items. We will discuss the following three different situations: (1) different coherence measurements, (2) different types of noises, (3) different database sizes. At the same time, the above discussions will consider the noise level varies from $0$ to $30\%$ i.e. $\eta=1$ to $\eta=0.7$, where $\eta=1$ corresponding to the ideal GSA.

\subsection{Different coherence measurements}\label{SubS41}
\begin{figure}\centering
\subfigure[$P_s(t)$]{\label{fig:subfig:a}\includegraphics[width=0.22\textwidth]{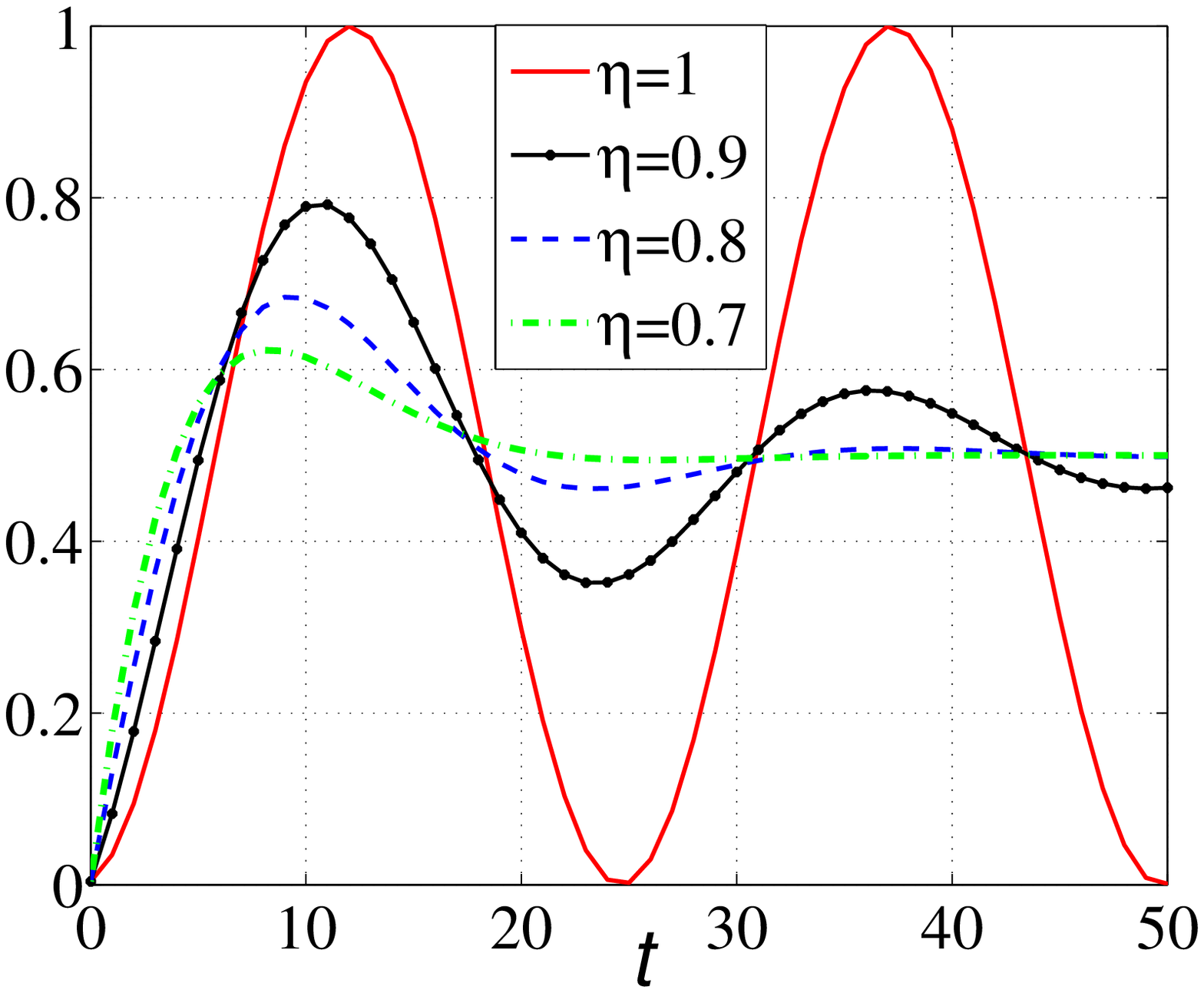}}\quad
\subfigure[$C_e(t)$]{\label{fig:subfig:b}\includegraphics[width=0.22\textwidth]{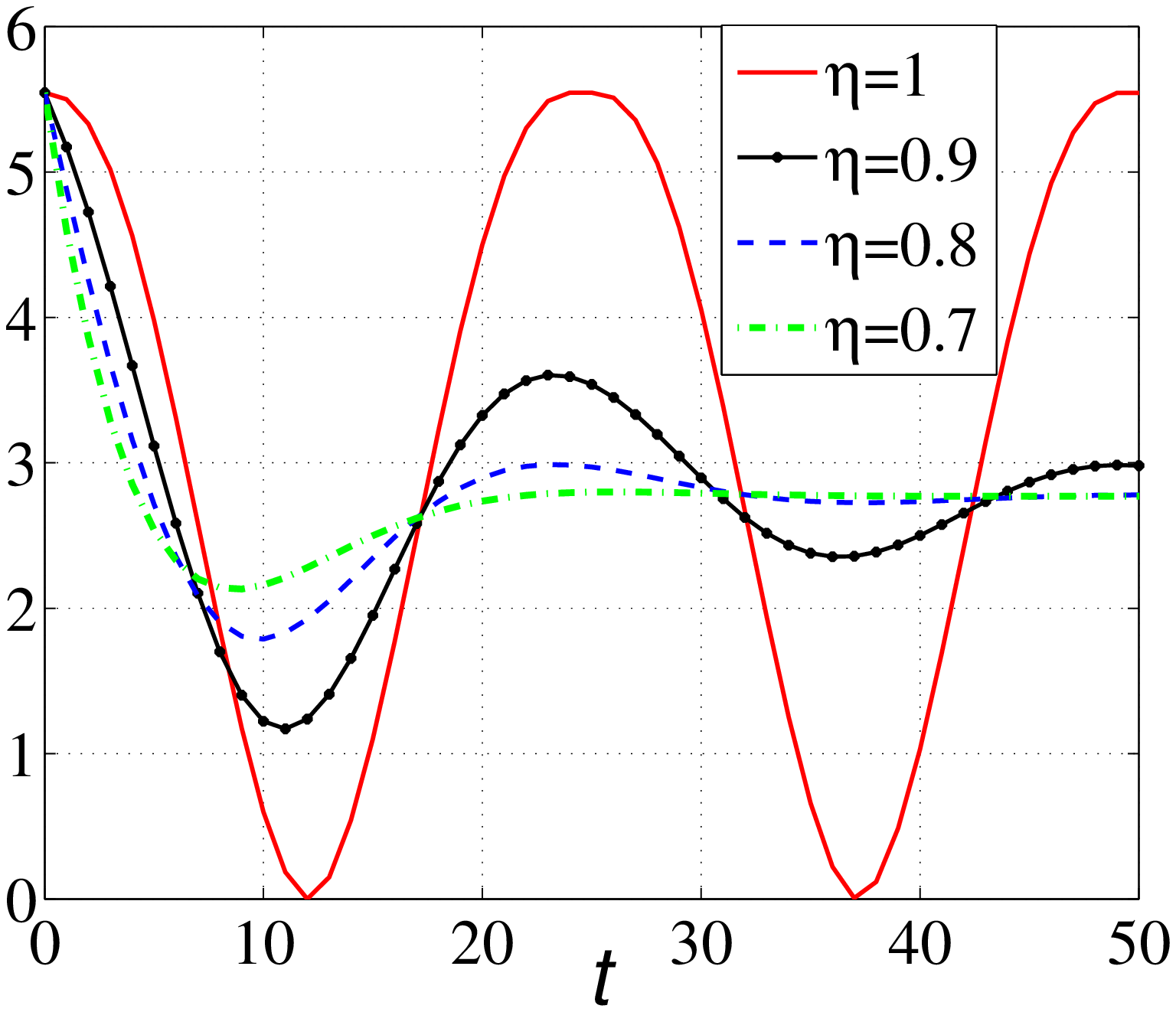}}\quad\\
\subfigure[$C_l(t)$]{\label{fig:subfig:c}
\includegraphics[width=0.22\textwidth]{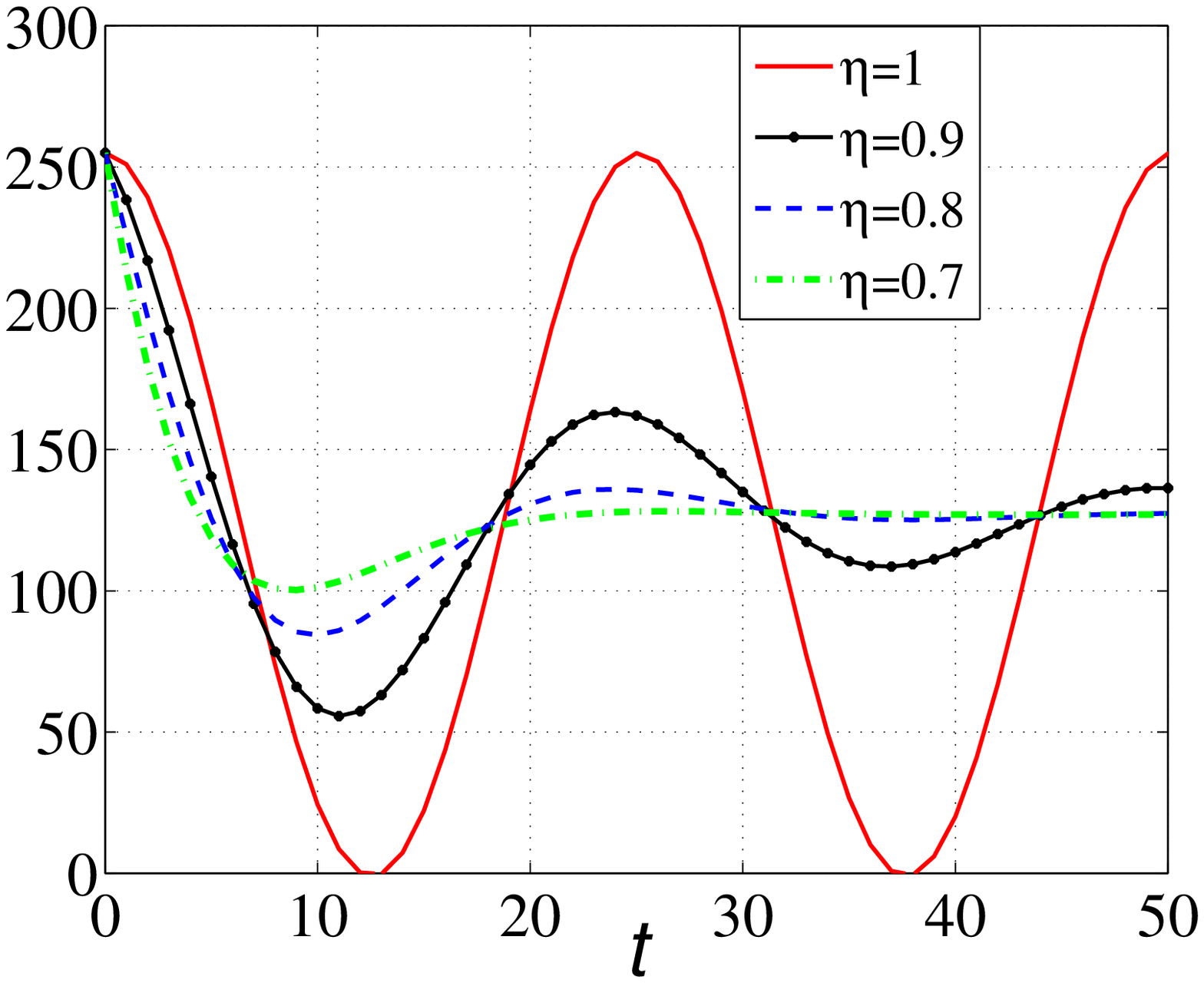}}\quad
\subfigure[$P_s \& 1-\mathds{N}(C_e) \& 1-\mathds{N}(C_l)$]{\label{fig:subfig:d}
\includegraphics[width=0.22\textwidth]{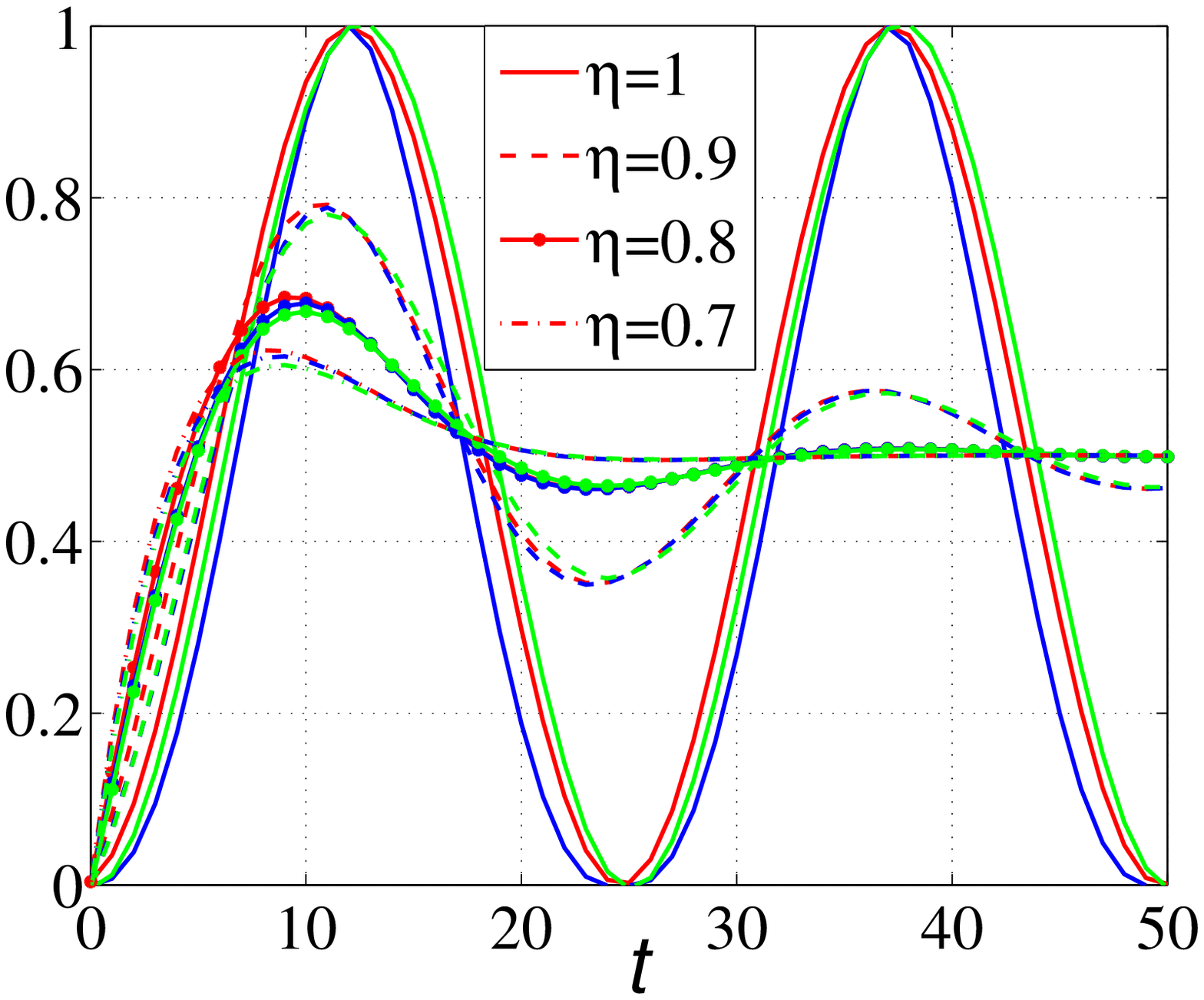}}\quad
\caption{\label{fig:B}(Color online) The success probability and Coherence vs. the number of iterations $t$ with Bit flip noise for $N=2^8$.
(For sub-figure \ref{fig:subfig:d}, $P_s(t)$ is red, $1-\mathds{N}(C_e(t))$ is blue and $1-\mathds{N}(C_l(t))$ is green with the same type of lines for the same noise level.)}
\end{figure}

Taking GSA with the bit flip noise as an example, we demonstrate how the success probability and coherence change with the number $t$ of search iterations in different measurements in Fig.\ref{fig:B}.
Note that to clear the trend of the bit flip noise affected on GSA, here we consider $t\in [0,4T]$.
The success probability $P_s(t)$ and the relative entropy of coherence $C_e(t)$ and $l_1$ norm of coherence $C_l(t)$ are illustrated in the sub-figures Fig.\ref{fig:subfig:a} to Fig.\ref{fig:subfig:c}.
Comparing with these sub-figures, we can observe that $P_s(t)$ always opposites to the coherence which is no matter measured by $C_e(\rho(t))$ or $C_l(t)$. At the same time, they are periodic oscillating decreasing except for the ideal case ($\eta=1$). Moreover, the more the noises are, the quicker they decrease.
Here we can clearly see their trends, but it is difficult to compare them uniformly.

\begin{figure}\centering
\subfigure[For Phase flip]
{\includegraphics[width=0.22\textwidth]{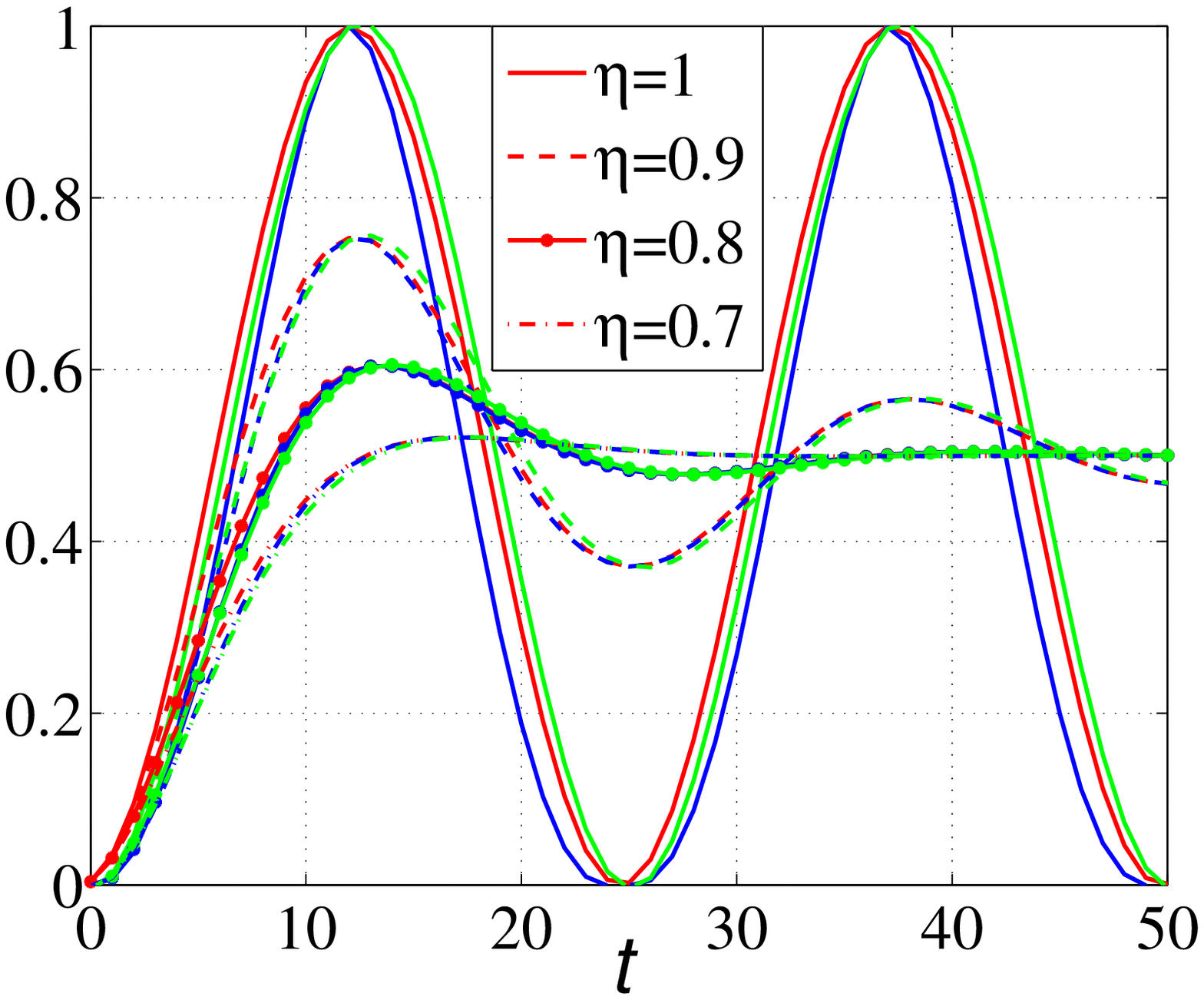}}
\subfigure[For Bit-Phase flip]
{\includegraphics[width=0.22\textwidth]{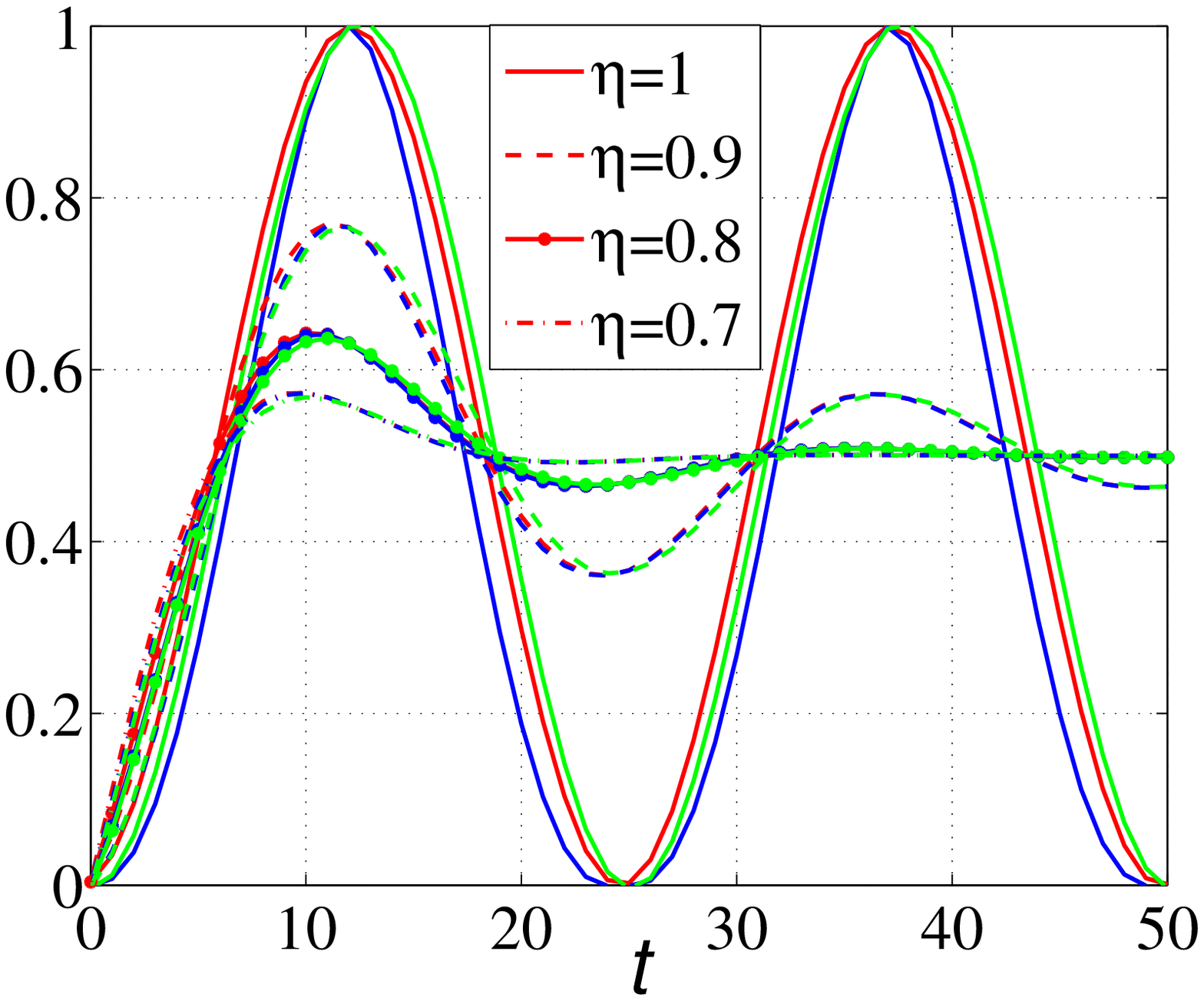}}
\caption{\label{fig:PBP}(Color online) The success probability and its complementary coherence vs. the number of iterations $t$ for $N=2^8$. (For the same noise level, $P_s(t)$ is red, $C_e(t)$ is blue and $C_l(t)$ is green with the same type of lines.)}
\end{figure}

For a better comparison, the success probability togethers with the normalization of the relative entropy of coherence and $l_1$ norm of coherence, i.e. $P_s(t)$ with $1-\mathds{N}(C_e(t))$ and $1-\mathds{N}(C_l(t))$, are presented in Fig.\ref{fig:subfig:d}. In that figure, they are plotted by  different type of lines for different noise levels. 
For the same noise level, they are plotted in different colors that $P_s(t)$ is red and $1-\mathds{N}(C_e(t))$ is blue and $1-\mathds{N}(C_l(t))$ is green.
We can see clearly that these three color lines are very close to each others.
As the number of iterations $t$ increases, they are closer and even overlap except for the ideal GSA with $\eta=1$.
At the same time, as the noise increases, they have the same phenomenon that are closer and even overlap at last.

According to the above discussion, we obtain that the success probability and the coherence, which are measured by the normalization of relative entropy of coherence and $l_1$ norm of coherence, satisfy the complementary relationship $P_s(t)\approx 1-\mathds{N}(C(t))$. Note that it doesn't matter if it is ideal or in noisy environments.

\subsection{Different types of noises}\label{SubS42}
We have discussed the GSA suffering from the bit flip noise in detail, now we consider other kinds of noises.
Since the normalization of coherence can measure the coherence better and more conveniently, in the following tests, we just discuss the normalization of relative entropy of coherence and $l_1$ norm of coherence.

As shown in Fig.\ref{fig:PBP}, it demonstrates $P_s(t)$ and $1-\mathds{N}(C_e(t))$ as well as $1-\mathds{N}(C_l(t))$ in GSA with the phase flip and bit-phase flip noise for $N=2^8$ and $t\in[0,4T]$.
We can observe a similar phenomenon that their values are very close for the same noise level. It means that the success probability and the coherence, which are measured by the normalization of relative entropy of coherence and $l_1$ norm of coherence, satisfy the complementary relationship $P_s(t)\approx 1-\mathds{N}(C(t))$.

\begin{figure}\centering
\subfigure[$N=2^8$]{\includegraphics[width=0.22\textwidth]{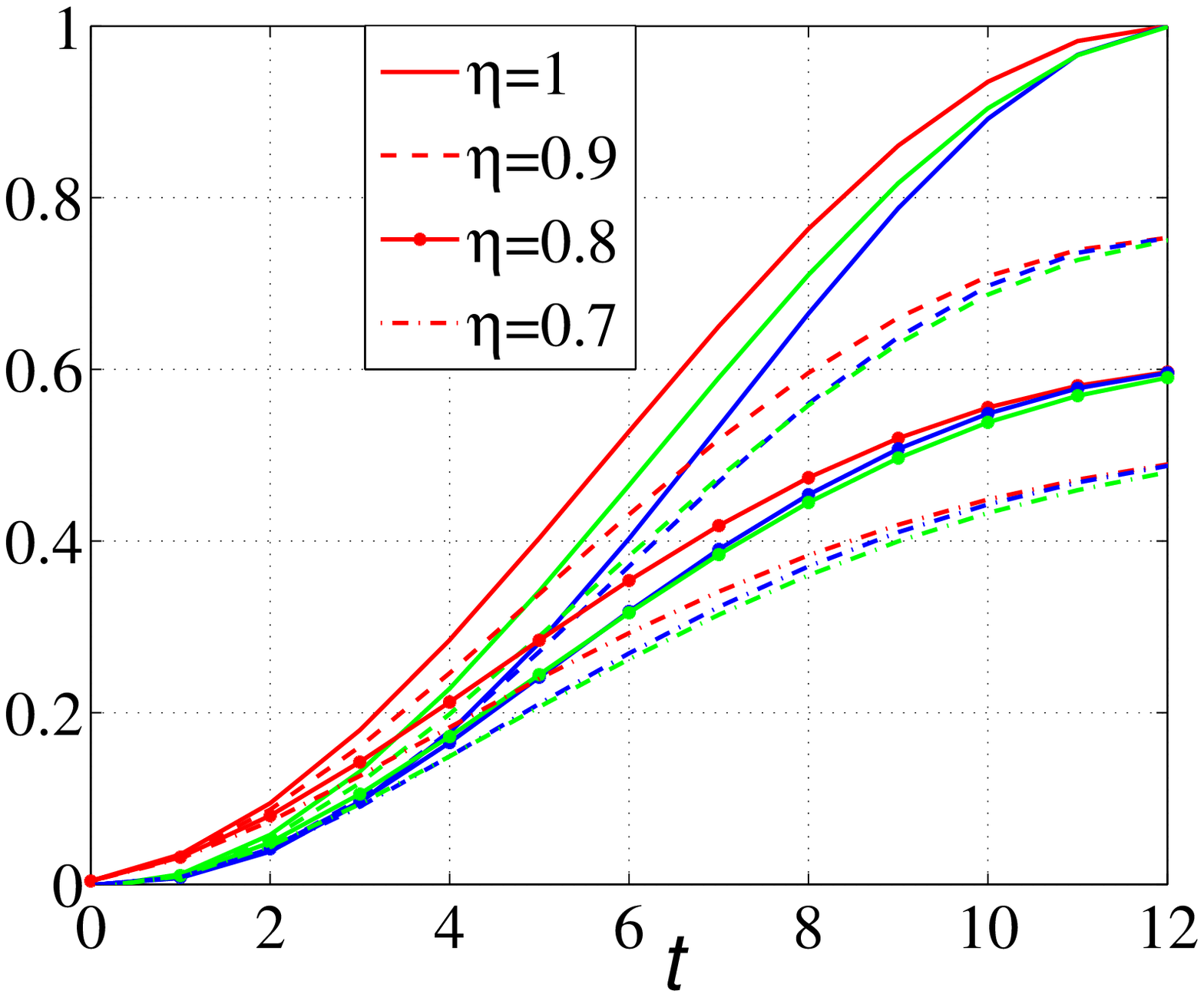}}
\subfigure[$N=2^{10}$]{\includegraphics[width=0.22\textwidth]{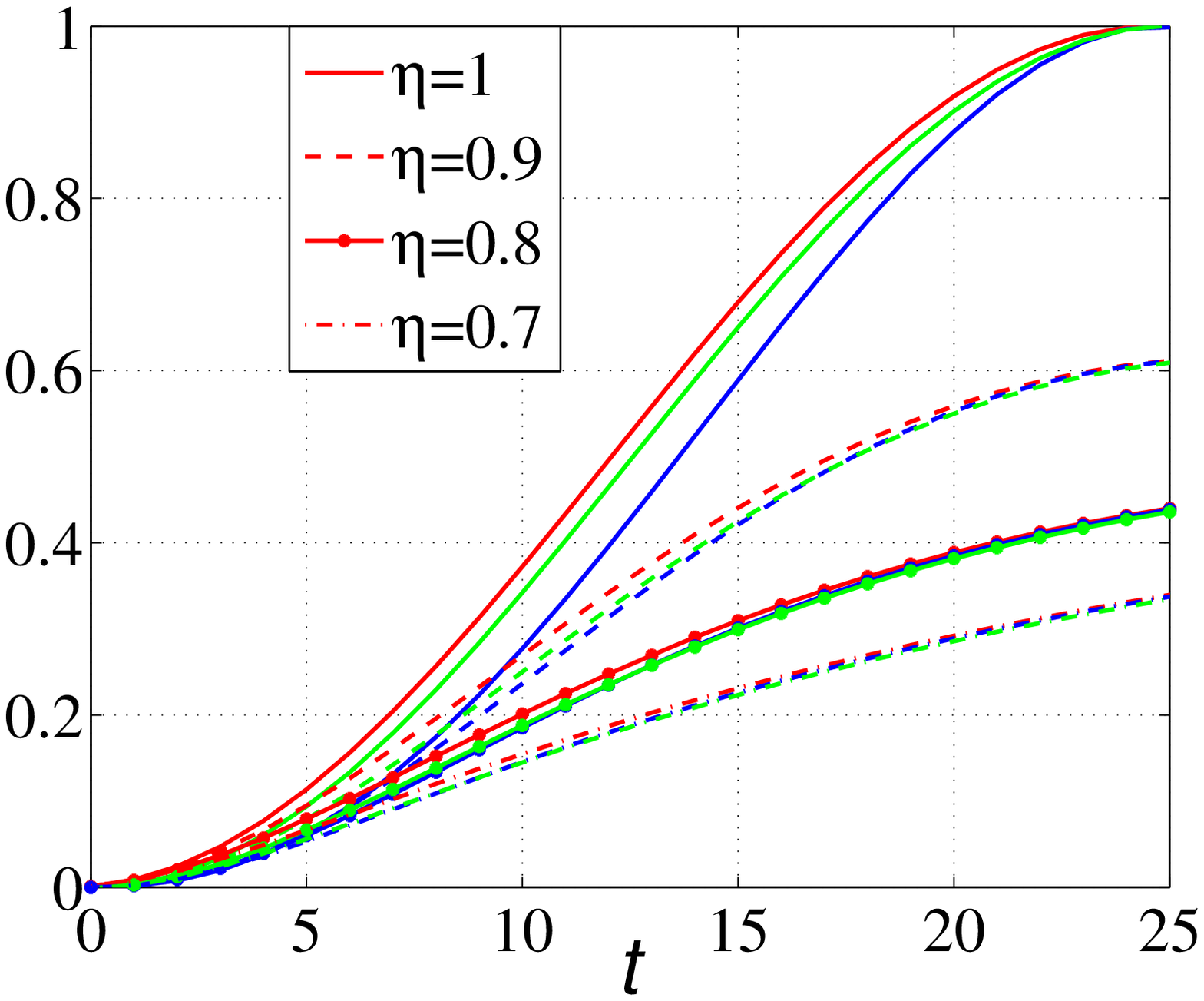}}
\subfigure[$N=2^{12}$]{\includegraphics[width=0.22\textwidth]{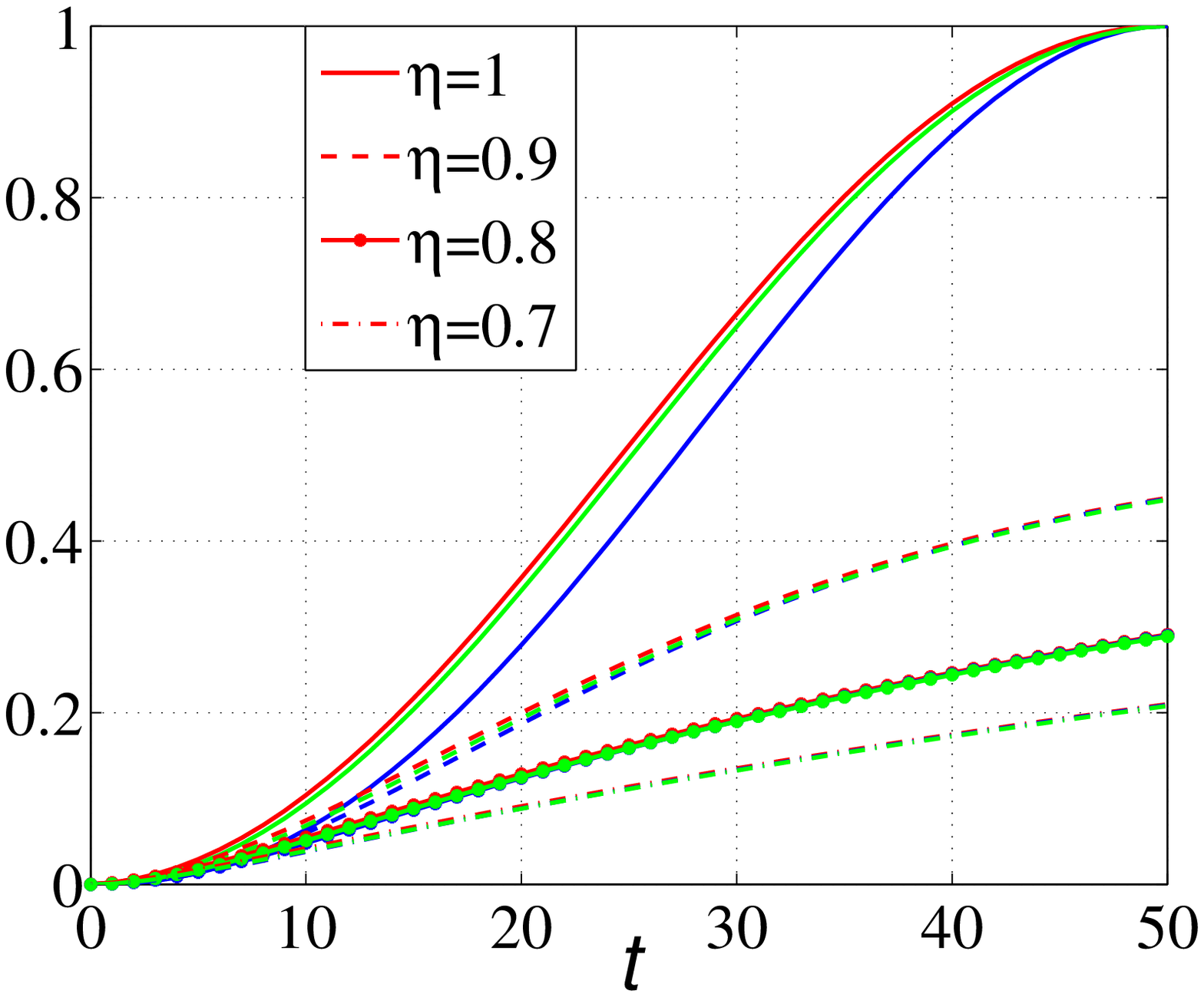}}
\subfigure[$N=2^{16}$]{\includegraphics[width=0.22\textwidth]{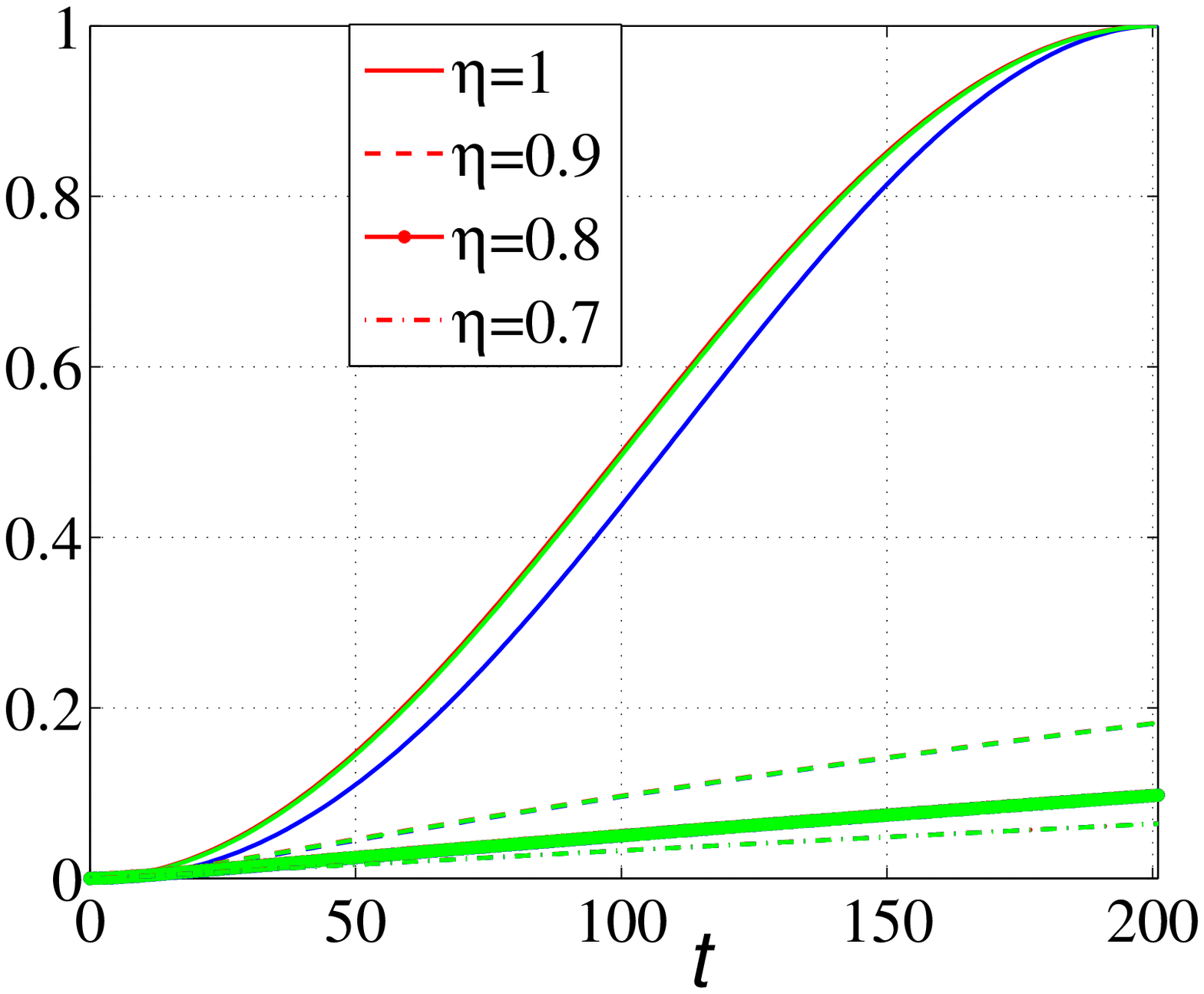}}
\caption{\label{fig:PSCT}(Color online) The success probability and its complementary coherence vs. the number of iterations $t$ with Phase flip noise. (For the same noise level, $P_s(t)$ is red, $1-\mathds{N}(C_e(t))$ is blue and $1-\mathds{N}(C_l(t))$ is green with the same type of lines.)}
\end{figure}

\subsection{Different database sizes}\label{SubS43}
As discuss in the above two subsections 
$P_s(t)$ and $1-\mathds{N}(C_e(t))$ and $1-\mathds{N}(C_l(t))$ in GSA tend to same as $t$ increases. To learn the difference of them, we will focus on the discussion of the number of iterations $t$ for $0<t\leq T$.
Consider the different database sizes $N=2^8,2^{10},2^{12},2^{16}$ and take the phase flip as example, we demonstrate the dynamics of them in Fig.\ref{fig:PSCT}.

As shown in Fig. \ref{fig:PSCT}, there are a little differences among $P_s(t), (1-\mathds{N}(C_e(t)))$ and $(1-\mathds{N}(C_l(t)))$ at first, and then the differences become smaller as $t$ increase and become the same as $t=T$.
At the same time, the differences among them become smaller as the noise increases ($\eta$ decreases).
In other words, there is a slight disparity in the complementarity between the success probability and the normalization coherence, however the disparity will disappear quickly as the noise or the database size increase.
The maximum even become smaller than $0.5$ as $N$ increases or the noise increases. To a certain extent, the quantum advantage will vanish as discussed in Ref.\cite{Pan21}.
Nevertheless, the complementary between the success probability and the normalization of coherence is still always holds.


\section{Discussion}\label{Sec5}
We had shown that $P_s(t)+\mathds{N}(C(t))\simeq 1$ which exhibits the complementarity between $P_s(t)$ and $\mathds{N}(C(t))$ for $M\leq N$, in terms of the relative entropy and the $l_1$ norm of coherence.
It holds in different environments including ideal and with noises. For intuition and visualization, we discuss the noise cases of bit flip (BF) and phase flip (PF) and bit-phase flip (BPF) in detail.
Considering the qubits are limited in the recent NISQ era, numerical experiments were presented for various limited database sizes and noise levels in different environments.

\begin{table}
\caption{The comparisons of some relevant works }\label{Tab1}
\resizebox{0.45\textwidth}{!}{
\begin{tabular}	{|c|c|c|c|c|c|}
\hline
Index &Shi\cite{Shi} & Rastegin\cite{Rastegin} & Pan\cite{Pan191} &Rastegin\cite{Rastegin21}&This work \\ \hline
Ideal & $\surd$ & $\surd$ & $\surd$ & $\surd$ & $\surd$\\ \hline
Noise & $\times$ &PF(PD) & $\times$ & AD & ..., PF(PD), BF, BPF \\ \hline
Coherence measurement & $C_e, C_l$ & $C_e$ & $ C_l$ & $C_e$ & $C_e, C_l$ \\ \hline
Normalization & $\times$ & $\times$ & $\surd$ & $\times$ & $\surd$ \\ \hline
The complexity of the relations & middle & complex & middle & complex & simple \\ \hline
\end{tabular}
}
\end{table}

Taking the previous results on the coherence of GSA into account, and to show the contribution of our results, the comparison with the most relevant previous works are listed in Table \ref{Tab1}.
We can observe clearly that our new results not only generalize these previous works, the most meaningful thing is that we present the main result in a simple and concise complementary form which unifies those previous works as shown in Table \ref{Tab1}.

\section{Conclusion}\label{Sec6}
Coherence is critical in quantum computing. Considering the current noisy quantum era and the important role of coherence in quantum algorithms, we discussed the relationship between the success probability $P_s(t)$ and the coherence of Grover search algorithm (GSA) in different environments.
At first, we defined the normalization coherence $\mathds{N}(C)$, where $C$ is a coherence measurement.
Then, we considered the coherence dynamics with the number of search iterations $t$ in GSA. We proved that the complementary between $P_s(t)$ and the coherence $\mathds{N}(C(t))$ for large $N$, in terms of the relative entropy and the $l_1$ norm of coherence.
In a simple and elegant way, we had shown that $P_s(t)+\mathds{N}(C(t))\simeq 1$.

At the same time, the numeral results showed that the complementary between $P_s(t)$ and $\mathds{N}(C(t))$ almost always holds, besides a slight difference for some parts of the process in the case of small $N$ or the ideal case. The larger the size $N$ is, the closer the success probability $P_s(t)$ is to its complementary coherence $1-\mathds{N}(C(t))$.

The complementarity between success rate and coherence is very meaningful, because it provides an idea to improve the success probability by manipulating its complementary coherence, and vice versa.
Therefore, this will benefit to quantum algorithms that operate in quantum computers especially in the NISQ era. It is noted that the complementarity is obtained within Bloch representation. Whether it holds or not considering in the whole $N$ dimensions is worth further research.

\acknowledgments
M. Pan was supported by Guangxi Natural Science Foundation (No.2019GXNSFBA245087), Guangxi Science and Technology Program (No.GuiKeAD21075020), Guangxi Key Laboratory of Cryptography and Information Security (No.202133).
S. Zheng was supported by the Major Key Project of PCL.


\bibliographystyle{eplbib}
\bibliography{CPCbib}

\end{document}


\maketitle

\tableofcontents

\clearpage

\section {Competitive Exclusion Principle (CEP)}
\subsection{The content of the CEP}
The notion of the competitive exclusion dates back to Charles Darwin’s theory of evolution, survival of the fittest. The earliest form of the CEP \cite{gause1934struggle, hardin1960competitive}, also referred as the Gause’s law, states that two species competing for the same limited resources cannot coexist at constant population densities. In the 1960s, MacArthur and Levin extended this principle to a generic case with an arbitrary number of resources \cite{macarthur1964competition, levin1970community}. The principle states that, in a well-mixed system of $M$ types of consumers feeding on $N$ types of resources, the number of consumer species in coexistence cannot exceeds that of the resources at steady state, i.e., $M \leq N$. (See the SI in Ref. \cite{wang2020overcome} for more details.)

\subsection{ The classical proof of the CEP}
In a classical paper \cite{macarthur1964competition, levin1970community}, MacArthur and Levin proposed a mathematical proof the CEP. We rephrase the idea of the proof in a simple case of $M=2$ and $N=1$, i.e., two consumer species $C_{1}$ and $C_{2}$ competing for one resource species $R$. In fact, it is easy to generalize this proof to higher dimensions with several types of consumers and resources. Then, the population dynamics of the species can be described as follows:

\begin{equation}
    \begin{cases}
    \dot{C_{i}}=C_{i}(f_{i}(R)-D_{i}),\  i=1,2; \\
    \dot{R}=g(R,C_{1},C_{2}). \\
    \end{cases}
\label{CEPequ}
\end{equation}
Here $C_{i}$ and $R$ represent the population abundances of consumers and resources, while the functional forms of $f_{i}(R)$, $g(R,C_{1},C_{2})$ are unspecific. $D_{i}$ stands for the mortality rate of the species $C_{i}$. If all consumer species can coexist at steady state, then $f_{i}(R)/D_{i}=1(i=1,2)$. In a 2-D representation, this requires that three lines, $y=f_{i}(R)/D_{i}=1(i=1,2)$ and $y=1$, share a common point, which is commonly impossible unless the model parameters satisfy special constraint (sets of Lebesgue measure zero). In a 3-D representation, the two planes, which correspond to , are parallel to each other, and hence do not share a common point (See also Fig. \ref{S5}a-b. See Ref. \cite{wang2020overcome} for details).

\section{Comparison of the functional response with Beddington-DeAngelis’ model in scenarios involving different types of pairwise encounters}\label{beddington}

\subsection{Beddington’s model}
In the 1970s, Beddington \cite{beddington1975mutual} proposed a mathematical model to describe the influence of predator interference on the functional response, where he applied handwaving derivations in a simple system with one type of consumers and one type of resources. In the same year, DeAngelis \cite{deangelis1975model} considered a related question and put forward a similar model. Essentially, both models are phenomenological, and they were called Beddington-DeAngelis model (B-D model) in the subsequent studies. In practice, the B-D model can be extended into scenarios involving different types of pairwise encounters with Beddington’s modelling method. In this section, we systematically compare the functional response in our mechanistic model with that of the B-D model in all the relevant scenarios.

Recalling Beddington’s analysis \cite{beddington1975mutual}, he considered a simple case of one consumer species $C$ and one resource species $R(M = 1, N = 1)$. In a well-mixed environment, an individual consumer meets a resource with rate $a$, while encounters another consumer with rate $a^{'}$. There are two other phenomenological parameters in this model, namely, the handling time $t_{h}$ and the wasting time $t_{w}$, which actually can be both determined by specifying the scenario and with further statistical physics modeling analysis. In fact, Beddington analyzed the searching efficiency $\Xi_{\text{B-D}}$ rather than the functional response $\mathcal{F}_{\text{B-D}}$, while both can be reciprocally derived with $\Xi_{\text{B-D}} \equiv \mathcal{F}_{\text{B-D}}/R$. Here $R$ stands for the population abundance of the resources, and the specific form of $\Xi_{\text{B-D}}$ is \cite{beddington1975mutual}:

\begin{equation}
    \Xi_{\text{B-D}}(R,C)=\frac{a}{1+at_{h}R+a^{'}t_{w}C^{'}},
\label{searchingequ}
\end{equation}
where $C^{'}=C-1$, with $C$ stands for the population abundance of the consumes. Generally, $C \gg 1$, and thus $C^{'} \approx C$.

\subsection{Scenario involving only chasing pair}
Here we consider the scenario involving only chasing pair for the simple case of one type of consumers and one type of resources ($M = 1, N = 1$). When an individual consumer is chasing a resource, they form a chasing pair.
$$\ce{$C$^{(F)} + $R$^{(F)} <=>[$a$][$d$] $C$^{(P)} \vee $R$^{(P)} ->T[$k$] $C$^{(F)}(+)}$$
where the superscript “F” stands for populations that are freely wandering, and “(+)” signifies gaining biomass (we count $C^{\text{(F)}}(+)$ as $C^{\text{(F)}}$). $C^{\text{(P)}} \vee R^{\text{(P)}}$ represents chasing pair (where “P” signifies pair), denoted as $x$. $a$, $d$ and $k$ stand for encounter rate, escape rate and capture rate, respectively. Hence, the total number of consumers and resources are $C \equiv C^{\text{(F)}} + x$ and $R \equiv R^{\text{(F)}} + x$. Then, the population dynamics of the consumers and resources follows:

\begin{equation}
    \begin{cases}
    \dot{x}=aC^{\text{(F)}}R^{\text{(F)}}-(k+d)x, \\
    \dot{C}=wkx-DC, \\
    \dot{R}=g(R,x,C). \\
    \end{cases}
\label{traseequ}
\end{equation}
where the functional form of $g(R,x,C)$ is unspecific, $D$ and $w$ represent the mortality rate of the consumer species and biomass conversion ratio, respectively. Since the consumption process is much faster than the birth/death process, thus, in deriving the functional response, the consumption process is supposed to be in fast equilibrium (i.e., $\dot{x}=0$). Then, we can solve for $x$ with:

\begin{equation}
x^{2}-(R+C+K)x+RC=0,
\label{xequ}
\end{equation}
where $K=\frac{k+d}{a}$, and then,

\begin{equation}
x=\frac{2RC}{(R+C+K)} \frac{1}{(1+\sqrt{1-\frac{4RC}{(R+C+K)^{2}}})}.
\label{xsolution}
\end{equation}
By definition, the functional response and search efficiency are:

\begin{subequations}
\begin{align}
& \mathcal{F}_{\text{CP}}(R,C)=\frac{kx}{C}, \\
& \Xi_{\text{CP}}(R,C)=\frac{kx}{RC}.
\end{align}
\label{FXiCP}
\end{subequations}
Hence, we obtain the functional response and search efficiency in this chasing-pair scenario:

\begin{subequations}
\begin{align}
& \mathcal{F}_{\text{CP}}(R,C)_{1}=k\frac{(R+C+K)}{2C} \left( 1-\sqrt{1-\frac{4RC}{(R+C+K)^{2}}} \right), \\
& \Xi_{\text{CP}}(R,C)_{1}=k\frac{(R+C+K)}{2RC}\left(1-\sqrt{1-\frac{4RC}{(R+C+K)^{2}}} \right).
\end{align}
\label{FXiCP1}
\end{subequations}
Since $\frac{4RC}{(R+C+K)^{2}}<4 \frac{C}{R} \ll 1$, by applying first order approximations in Eq. \eqref{FXiCP1}, we obtain $\sqrt{1-\frac{4RC}{(R+C+K)^{2}}} \approx 1- \frac{2RC}{(R+C+K)^{2}}$. Then the functional response and search efficiency are:

\begin{subequations}
\begin{align}
& \mathcal{F}_{\text{CP}}(R,C)_{2}=k\frac{R}{R+C+K}, \\
& \Xi_{\text{CP}}(R,C)_{2}=\frac{k}{R+C+K}.
\end{align}
\label{FXiCP2}
\end{subequations}
Evidently, there is entirely no predator interference within the chasing-pair scenario, yet the functional response form is identical to the B-D model involving intraspecific interference (see Eq. \ref{searchingequ}). Meanwhile, by applying first order approximations in the denominator of Eq. \ref{xsolution}, we have $x \approx \frac{RC}{(R+C+K)-\frac{RC}{(R+C+K)}}$. Hence,

\begin{subequations}
\begin{align}
& \mathcal{F}_{\text{CP}}(R,C)_{3}=k\frac{R}{(R+C+K)-\frac{RC}{(R+C+K)}}, \\
& \Xi_{\text{CP}}(R,C)_{3}=\frac{k}{(R+C+K)-\frac{RC}{(R+C+K)}}.
\end{align}
\label{FXiCP3}
\end{subequations}
In the case that $R \gg C$, then $R \gg C>x=R-R^{\text{(F)}}$. By applying $R \approx R^{\text{(F)}}$ in Eq. \ref{traseequ}, we obtain $x \approx \frac{RC}{R+K}$. Then,

\begin{subequations}
\begin{align}
& \mathcal{F}_{\text{CP}}(R,C)_{4}=k\frac{R}{R+K}, \\
& \Xi_{\text{CP}}(R,C)_{4}=\frac{k}{R+K}.
\end{align}
\label{FXiCP4}
\end{subequations}
To compare these functional responses with that of the B-D model, we determine the parameters $t_{h}$ and $t_{w}$ in the B-D model by calculating their average value in a stochastic framework. Then $\langle t_{h} \rangle=\frac{1}{k}$ and $\langle t_{w} \rangle=\frac{1}{d^{'}}$(in the chasing-pair scenario, $a^{'}=0$), and thus,

\begin{subequations}
\begin{align}
& \Xi_{\text{B-D}}(R,C)=\frac{a}{1+Ra/k}=\frac{k}{k/a+R}, \\
& \mathcal{F}_{\text{B-D}}(R,C)=\frac{kR}{k/a+R}.
\end{align}
\label{FXiBD}
\end{subequations}
In the special case of $d=0$ and $R \gg C$, $\Xi_{\text{B-D}}(R,C)=\Xi_{\text{CP}}(R,C)_{4}$, the B-D model is consistent with our mechanistic model. However, the discrepancy can be large out of this region (e.g., $d \gg 0$) (Fig. \ref{S2}).

\subsection{Scenario involving chasing pair and intraspecific interference}\label{beddingtonchasintra}

Here we consider the scenario with additional involvement of intraspecific interference in the simple case of $M=1$ and $N=1$:
$$\ce{$C$^{(F)} + $R$^{(F)} <=>[$a$][$d$] $C$^{(P)} \vee $R$^{(P)} ->T[$k$] $C$^{(F)}(+)}$$
$$\ce{$C$^{(F)} + $C$^{(F)} <=>[$a^{'}$][$d^{'}$] $C$^{(P)} \vee $C$^{(P)}}$$
where $C^{\text{(P)}} \vee C^{\text{(P)}}$ stands for the intraspecific predator interference pair, denoted as $y$. $a^{'}$ and $d^{'}$ represent the encounter rate and separation rate of the interference pair, respectively. Then, the total population of consumers and resources are $C \equiv C^{\text{(F)}}+x+2y$ and $R \equiv R^{\text{(F)}}+x$. Hence the population dynamics of the consumers and resources can be described as follows:

\begin{equation}
    \begin{cases}
    \dot{x}=aC^{\text{(F)}}R^{\text{(F)}}-(k+d)x, \\
    \dot{y}=a^{'}[C^{\text{(F)}}]^{2}-d^{'}y, \\
    \dot{C}=wkx-DC, \\
    \dot{R}=g(R,x,C). \\
    \end{cases}
\label{intrachaseequ}
\end{equation}
The consumption process and interference process are supposed to be in fast equilibrium (i.e., $\dot{x}=0, \dot{y}=0$), then we can solve for $x$ with:

\begin{equation}
x^{3}+ \phi_{2}x^{2}+ \phi_{1}x+ \phi_{0}=0,
\label{xintertrasequ}
\end{equation}
where $\phi_{0} \approx -CR^{2},\phi_{1}=2CR+KR+R^{2},\phi_{2}=2 \beta K^{2}-K-C-2R$, with $\beta=a^{'}/d^{'}$. The discriminant of Eq. \ref{xintertrasequ} (denoted as $\Lambda$) is

\begin{equation}
\Lambda=-4\psi^{3}-27\varphi^{2},
\label{lambdadiscriminant}
\end{equation}
with $\psi=\phi_{1}-(\phi_{2})^{2}/3$ and $\varphi=\phi_{0}-\phi_{1}\phi_{2}/3+2(\phi_{2})^{3}/27$. When $\Lambda<0$, there are one real solution $x_{S1}$ and two complex solutions $x_{S2}, x_{S3}$, which are

\begin{equation}
x_{S1}=\theta_{1}+\theta_{2}-\phi_{2}/3, x_{S2}=\omega\theta_{1}+\omega^{2}\theta_{2}-\phi_{2}/3, x_{S1}=\omega^{2}\theta_{1}+\omega\theta_{2}-\phi_{2}/3,
\label{x123solution}
\end{equation}
where $\omega=-1/2+i\sqrt{3}/2$ ($i$ stands for the imaginary unit), $\theta_{1}=(-\varphi/2+\sqrt{-Lambda/108})^{1/3}$, and $\theta_{2}=(-\varphi/2-\sqrt{-Lambda/108})^{1/3}$. On the other hand, when $Lambda>0$, there are three real 
solutions $x_{S1}, x_{S2}$, and $x_{S3}$, which are

\begin{equation}
x_{S1}=\psi^{'}cos\varphi^{'}-\phi_{2}/3, x_{S2}=\psi^{'}cos(\varphi^{'}+\frac{2\pi}{3})-\phi_{2}/3, x_{S1}=\psi^{'}cos(\varphi^{'}+\frac{4\pi}{3})-\phi_{2}/3,
\label{xS123solution}
\end{equation}
where $\psi^{'}=(-4\psi/3)^{1/2}$, and $\varphi^{'}=arccos(-(-\psi/3)^{-3/2}\varphi/2)/3$. Note that $x \in [0,min(R,C)]$, then we obtain the feasible solution of $x$(exact solution), and thus, the functional response and search efficiency are

\begin{subequations}
\begin{align}
& \mathcal{F}_{\text{A}}(R,C)_{1}=\frac{kx}{C}, \\
& \Xi_{\text{A}}(R,C)_{1}=\frac{kx}{RC}.
\end{align}
\label{FXiA1}
\end{subequations}
In the case that $R \gg C$, then $R \gg C>x$, and thus $R^{\text{(F)}} \approx R$. The consumption process is supposed to be in fast equilibrium (i.e., $\dot{x}=0,\dot{y}=0$), then we obtain

\begin{equation}
x \approx \frac{RC}{\sqrt{[\frac{1}{2}(K+R)]^{2}+2C\beta K^{2}}+\frac{1}{2}(K+R)},
\label{xappsolution}
\end{equation}
and thus,

\begin{subequations}
\begin{align}
& \mathcal{F}_{\text{A}}(R,C)_{2}=k\frac{R}{\sqrt{[\frac{1}{2}(K+R)]^{2}+2C\beta K^{2}}+\frac{1}{2}(K+R)}, \\
& \Xi_{\text{A}}(R,C)_{2}=k\frac{1}{\sqrt{[\frac{1}{2}(K+R)]^{2}+2C\beta K^{2}}+\frac{1}{2}(K+R)}.
\end{align}
\label{FXiA2}
\end{subequations}
When $\beta \ll \frac{1}{8C}$ or $8\beta C/(1+R/K)^{2} \ll 1$, by applying first order approximations to the denominator of Eq. \ref{xappsolution}, we have

\begin{equation}
x \approx \frac{RC}{(K+R)+\frac{2K}{(1+R/K)}\beta C},
\label{xappll1}
\end{equation}
and then,

\begin{subequations}
\begin{align}
& \mathcal{F}_{\text{A}}(R,C)_{3}=k\frac{R}{(K+R)+\frac{2K}{(1+R/K)}\beta C}, \\
& \Xi_{\text{A}}(R,C)_{3}=k\frac{1}{(K+R)+\frac{2K}{(1+R/K)}\beta C}.
\end{align}
\label{FXiA3}
\end{subequations}
In the case that $8\beta C/(1+R/K)^{2} \gg 1$, with first order approximations, we obtain

\begin{equation}
x \approx \frac{RC}{K\sqrt{2C\beta}+\frac{(K+R)^{2}}{8K\sqrt{2C\beta}}+\frac{1}{2}(K+R)},
\label{xappgg1}
\end{equation}
and thus,

\begin{subequations}
\begin{align}
& \mathcal{F}_{\text{A}}(R,C)_{4}=k\frac{R}{K\sqrt{2C\beta}+\frac{(K+R)^{2}}{8K\sqrt{2C\beta}}+\frac{1}{2}(K+R)}, \\
& \Xi_{\text{A}}(R,C)_{4}=k\frac{1}{K\sqrt{2C\beta}+\frac{(K+R)^{2}}{8K\sqrt{2C\beta}}+\frac{1}{2}(K+R)}.
\end{align}
\label{FXiA4}
\end{subequations}
Meanwhile, the B-D model can only match to the case with $d = 0$. By calculating the average values of $t_{h}$ and $t_{w}$ in a stochastic framework, and then $\langle t_{h} \rangle=\frac{1}{k}, \langle t_{w} \rangle=\frac{1}{d^{'}}$. Thus

\begin{subequations}
\begin{align}
& \Xi_{\text{A}}^{\text{B-D}}(R,C)=\frac{a}{1+\frac{a}{k}R+\frac{a^{'}}{d^{'}}C}=\frac{a}{1+R/K \mid_{d=0}+\beta C}, \\
& \mathcal{F}_{\text{A}}^{\text{B-D}}(R,C)=\frac{aR}{1+R/K \mid_{d=0}+\beta C}.
\end{align}
\label{FXiABD}
\end{subequations}
In fact, the searching efficiency (and thus the functional response) of the B-D model do not match with either the rigorous form $\Xi_{\text{A}}(R,C)_{1}$, the quasi rigorous form $\Xi_{\text{A}}(R,C)_{2}$, or the more simplified forms $\Xi_{\text{A}}(R,C)_{3}$ and 
$\Xi_{\text{A}}(R,C)_{4}$. However, the discrepancy is small when $d \approx 0$ and $R \gg C$ (Fig. \ref{S3}). Intuitively, when $\beta \ll \frac{1}{8C}$
and $d=0$, then $\Xi_{\text{A}}(R,C)_{3}=\frac{a}{1+\frac{a}{k}R+\frac{2}{(1+R/K)}\beta C_{i}}$. Consequently, if $R/K_{i}=x/C_{i}^{\text{(F)}}<1$, then $\frac{2}{(1+R/K_{i})} \in [1,2]$. In this case, the difference between $\Xi_{\text{A}}^{\text{B-D}}$ and $\Xi_{\text{A}}(R,C)_{3}$ is small. Actually, the above analysis also applies to cases with more than one types of consumer species (i.e., for cases with $M >1$).

\subsection{Scenario involving chasing pair and interspecific interference}\label{sectionchasinter}

Next, we consider the scenario involving chasing pair and interspecific interference in the case of $M=2$ and $N=1$:

$$\ce{$C_{i}$^{(F)} + $R$^{(F)} <=>[$a_{i}$][$d_{i}$] $C_{i}$^{(P)} \vee $R$^{(P)} ->[$k_{i}$] $C_{i}$^{(F)}(+)},i=1,2$$
$$\ce{$C_{1}$^{(F)} + $C_{2}$^{(F)} <=>[$a^{'}_{12}$][$d^{'}_{12}$] $C_{1}$^{(P)} \vee $C_{2}$^{(P)}}$$
where $C_{1}^{\text{(P)}} \vee C_{2}^{\text{(P)}}$ stands for the interspecific interference pair, denoted as $z$. $a^{'}_{12}$ and $d^{'}_{12}$ represent the encounter rate and separation rate of the interference pair, respectively. Then, the total population of consumers and resources are $C_{i} \equiv C_{i}^{\text{(F)}}+x_{i}+z$ and $R \equiv R^{\text{(F)}}+x_{1}+x_{2}$. The population dynamics of the consumers and resources follows:

\begin{equation}
    \begin{cases}
    \dot{x_{i}}=a_{i}C_{i}^{\text{(F)}}R^{\text{(F)}}-(k_{i}+d_{i})x_{i},i=1,2; \\
    \dot{z}=a_{12}^{'}C_{1}^{\text{(F)}}C_{2}^{\text{(F)}}-d_{12}^{'}z, \\
    \dot{C_{i}}=w_{i}k_{i}x_{i}-D_{i}C_{i}, \\
    \dot{R}=g(R,x_{1},x_{2},C_{1},C_{2}). \\
    \end{cases}
\label{interchaseequ}
\end{equation}
where the functional form of $g(R,x_{1},x_{2},C_{1},C_{2})$ is unspecific, while $D_{i}$ and $w_{i}$ represents the mortality rates of the two consumers species and biomass conversion ratios. Still, the consumption/interference process is supposed to be in fast equilibrium, i.e., $\dot{x_{i}}=0,\dot{z}=0$. In the case that $R \gg C_{1}+C_{2}>x_{1}+x_{2}$, by applying $R^{\text{(F)}} \approx R$, we obtain

\begin{subequations}
\begin{align}
& x_{1} \approx \frac{2C_{1}(R/K_{2}+1)R/K_{1}}{\sqrt{[\gamma (C_{2}-C_{1})+(\frac{R}{K_{1}}+1)(\frac{R}{K_{2}}+1)]^{2}+4\gamma C_{1}(\frac{R}{K_{1}}+1)(\frac{R}{K_{2}}+1)}+\gamma (C_{2}-C_{1})+(\frac{R}{K_{1}}+1)(\frac{R}{K_{2}}+1)}, \\
& x_{2} \approx \frac{2C_{2}(R/K_{1}+1)R/K_{2}}{\sqrt{[\gamma (C_{1}-C_{2})+(\frac{R}{K_{1}}+1)(\frac{R}{K_{2}}+1)]^{2}+4\gamma C_{2}(\frac{R}{K_{1}}+1)(\frac{R}{K_{2}}+1)}+\gamma (C_{1}-C_{2})+(\frac{R}{K_{1}}+1)(\frac{R}{K_{2}}+1)}.
\end{align}
\label{xsol}
\end{subequations}
and then, the searching efficiencies and functional responses are:

\begin{subequations}
\begin{align}
& \Xi_{1}(R,C_{1},C_{2})_{1}=\frac{2k_{1}(R/K_{2}+1)/K_{1}}{\parbox{2.75in}{$\gamma (C_{2}-C_{1})+(\frac{R}{K_{1}}+1)(\frac{R}{K_{2}}+1)+\sqrt{[\gamma (C_{2}-C_{1})+(\frac{R}{K_{1}}+1)(\frac{R}{K_{2}}+1)]^{2}+4\gamma C_{1}(\frac{R}{K_{1}}+1)(\frac{R}{K_{2}}+1)}$}}, \\
& \Xi_{2}(R,C_{1},C_{2})_{1}=\frac{2k_{2}(R/K_{1}+1)/K_{2}}{\parbox{2.75in}{$\gamma (C_{1}-C_{2})+(\frac{R}{K_{1}}+1)(\frac{R}{K_{2}}+1)+\sqrt{[\gamma (C_{1}-C_{2})+(\frac{R}{K_{1}}+1)(\frac{R}{K_{2}}+1)]^{2}+4\gamma C_{2}(\frac{R}{K_{1}}+1)(\frac{R}{K_{2}}+1)}$}}, \\
& \mathcal{F}_{1}(R,C_{1},C_{2})_{1}=\frac{2k_{1}(R/K_{2}+1)R/K_{1}}{\parbox{2.75in}{$\gamma (C_{2}-C_{1})+(\frac{R}{K_{1}}+1)(\frac{R}{K_{2}}+1)+\sqrt{[\gamma (C_{2}-C_{1})+(\frac{R}{K_{1}}+1)(\frac{R}{K_{2}}+1)]^{2}+4\gamma C_{1}(\frac{R}{K_{1}}+1)(\frac{R}{K_{2}}+1)}$}}, \\
& \mathcal{F}_{2}(R,C_{1},C_{2})_{1}=\frac{2k_{2}(R/K_{1}+1)R/K_{2}}{\parbox{2.75in}{$\gamma (C_{1}-C_{2})+(\frac{R}{K_{1}}+1)(\frac{R}{K_{2}}+1)+\sqrt{[\gamma (C_{1}-C_{2})+(\frac{R}{K_{1}}+1)(\frac{R}{K_{2}}+1)]^{2}+4\gamma C_{2}(\frac{R}{K_{1}}+1)(\frac{R}{K_{2}}+1)}$}}.
\end{align}
\label{FXi1}
\end{subequations}
Since $\frac{4\gamma^{2}C_{1}C_{2}}{[\gamma C_{1} +\gamma C_{2})+(\frac{R}{K_{1}}+1)(\frac{R}{K_{2}}+1)]^{2}}<1$, by applying first order approximation to 
the denominator of Eq. \ref{xsol}, we obtain:

\begin{subequations}
\begin{align}
& x_{1} \approx \frac{C_{1}R}{(R+K_{1})+\frac{\gamma K_{1}K_{2}C_{2}}{(R+K_{2})}-\frac{\gamma^{2} K_{1}K_{2}C_{1}C_{2}}{[\gamma (C_{1}+C_{2})+(\frac{R}{K_{1}}+1)(\frac{R}{K_{2}}+1)](R+K_{2})}}, \\
& x_{2} \approx \frac{C_{2}R}{(R+K_{2})+\frac{\gamma K_{1}K_{2}C_{1}}{(R+K_{1})}-\frac{\gamma^{2} K_{1}K_{2}C_{1}C_{2}}{[\gamma (C_{1}+C_{2})+(\frac{R}{K_{1}}+1)(\frac{R}{K_{2}}+1)](R+K_{1})}}.
\end{align}
\label{xsolapp}
\end{subequations}
and the searching efficiencies and functional responses are

\begin{subequations}
\begin{align}
& \Xi_{1}(R,C_{1},C_{2})_{2}=\frac{k_{1}}{(R+K_{1})+\frac{\gamma K_{1}K_{2}C_{2}}{(R+K_{2})}-\frac{\gamma^{2} K_{1}K_{2}C_{1}C_{2}}{[\gamma (C_{1}+C_{2})+(\frac{R}{K_{1}}+1)(\frac{R}{K_{2}}+1)](R+K_{2})}}, \\
& \Xi_{2}(R,C_{1},C_{2})_{2}=\frac{k_{2}}{(R+K_{2})+\frac{\gamma K_{1}K_{2}C_{1}}{(R+K_{1})}-\frac{\gamma^{2} K_{1}K_{2}C_{1}C_{2}}{[\gamma (C_{1}+C_{2})+(\frac{R}{K_{1}}+1)(\frac{R}{K_{2}}+1)](R+K_{1})}}, \\
& \mathcal{F}_{1}(R,C_{1},C_{2})_{2}=\frac{k_{1}R}{(R+K_{1})+\frac{\gamma K_{1}K_{2}C_{2}}{(R+K_{2})}-\frac{\gamma^{2} K_{1}K_{2}C_{1}C_{2}}{[\gamma (C_{1}+C_{2})+(\frac{R}{K_{1}}+1)(\frac{R}{K_{2}}+1)](R+K_{2})}}, \\
& \mathcal{F}_{2}(R,C_{1},C_{2})_{2}=\frac{k_{2}R}{(R+K_{2})+\frac{\gamma K_{1}K_{2}C_{1}}{(R+K_{1})}-\frac{\gamma^{2} K_{1}K_{2}C_{1}C_{2}}{[\gamma (C_{1}+C_{2})+(\frac{R}{K_{1}}+1)(\frac{R}{K_{2}}+1)](R+K_{1})}}.
\end{align}
\label{FXi2}
\end{subequations}
Likewise, the B-D model can only match to cases with $d = 0$. By calculating the average values in a stochastic framework, we obtain $\langle t_{h}^{i} \rangle=\frac{1}{k_{i}}, \langle t_{w}^{i} \rangle=\frac{1}{d_{12}^{'}}$. Thus

\begin{subequations}
\begin{align}
& \Xi_{1}^{\text{B-D}}(R,C_{1},C_{2})=\frac{a_{1}}{1+\frac{a_{1}}{k_{1}}R+\frac{a_{12}^{'}}{d_{12}^{'}}C_{2}}=\frac{a_{1}}{1+R/K_{1} \mid_{d=0}+\gamma C_{2}}, \\
& \Xi_{2}^{\text{B-D}}(R,C_{1},C_{2})=\frac{a_{2}}{1+\frac{a_{2}}{k_{2}}R+\frac{a_{12}^{'}}{d_{12}^{'}}C_{1}}=\frac{a_{2}}{1+R/K_{2} \mid_{d=0}+\gamma C_{1}}.
\end{align}
\label{Xi12BD}
\end{subequations}
Consequently,

\begin{subequations}
\begin{align}
& \mathcal{F}_{1}^{\text{B-D}}(R,C_{1},C_{2})=\frac{a_{1}R}{1+R/K_{1} \mid_{d=0}+\gamma C_{2}}, \\
& \mathcal{F}_{2}^{\text{B-D}}(R,C_{1},C_{2})=\frac{a_{2}R}{1+R/K_{2} \mid_{d=0}+\gamma C_{1}}.
\end{align}
\label{F12ABD}
\end{subequations}
Evidently, the searching efficiency in the B-D model do not match with either the quasi rigorous form $\Xi_{i}(R,C_{1},C_{2})_{1}$, or the simplified form $\Xi_{i}(R,C_{1},C_{2})_{2}$. However, the discrepancy can be small when $d \approx 0$ and $R \gg C$ (Fig. \ref{S4}). Intuitively, when $\gamma \ll min(C_{1}^{-1},C_{2}^{-1})$, we have

\begin{subequations}
\begin{align}
& \Xi_{1}^{\text{B-D}}(R,C_{1},C_{2})_{2} \approx \frac{a_{1}}{(1+\frac{a_{1}}{k_{1}}R)+\frac{\gamma C_{2}}{R/K_{2}+1}}, \\
& \Xi_{2}^{\text{B-D}}(R,C_{1},C_{2})_{2} \approx \frac{a_{2}}{(1+\frac{a_{2}}{k_{2}}R)+\frac{\gamma C_{1}}{R/K_{1}+1}}.
\end{align}
\label{Xi12BD2}
\end{subequations}
Consequently, if $R/K_{i}=x/C_{i}^{\text{(F)}}<1$, and then $\frac{1}{(1+R/K_{i})} \in [0.5,1]$, Thus, in this case, the difference between $\Xi_{i}^{\text{B-D}}$ and $\Xi_{i}^{\text{B-D}}(R,C_{1},C_{2})_{2}$ is small.

\section{Scenario involving chasing pair and intraspecific interference}

\subsection{Two consumers species competing for one resource species} \label{2C1R}

We consider the scenario involving chasing pair and intraspecific interference in the 
simple case of $M=2$ and $N=1$ (Fig. \ref{S9}a-b):
$$\ce{$C_{i}$^{(F)} + $R$^{(F)} <=>[$a_{i}$][$d_{i}$] $C_{i}$^{(P)} \vee $R$^{(P)} ->[$k_{i}$] $C_{i}$^{(F)}(+)}$$
$$\ce{$C_{i}$^{(F)} + $C_{i}$^{(F)} <=>[$a_{i}^{'}$][$d_{i}^{'}$] $C_{i}$^{(P)} \vee $C_{i}$^{(P)}},i=1,2$$
Here, the variables and parameters are just extended from the case of $M=1$ and $N=1$ (see SI Sec. \ref{beddingtonchasintra}). The total number of consumers and resources are $C_{i} \equiv C_{i}^{\text{(F)}}+x_{i}+2y_{i}$ and $R \equiv R^{\text{(F)}}+\sum\limits_{i=1}^{2}x_{i}$. Hence, the population dynamics of the 
consumers and resources can be described as follows:

\begin{equation}
    \begin{cases}
    \dot{x_{i}}=a_{i}C_{i}^{\text{(F)}}R^{\text{(F)}}-(k_{i}+d_{i})x_{i},i=1,2; \\
    \dot{y_{i}}=a_{i}^{'}[C_{i}^{\text{(F)}}]^{2}-d_{i}^{'}y_{i}, \\
    \dot{C_{i}}=w_{i}k_{i}x_{i}-D_{i}C_{i}, \\
    \dot{R}=g(R,x_{1},x_{2},C_{1},C_{2}). \\
    \end{cases}
\label{intrachaseequ2}
\end{equation}
where the functional form of $g(R,x_{1},x_{2},C_{1},C_{2})$ is unspecified. For simplicity, we define $K_{i} \equiv (d_{i}+k_{i})/a_{i}, \alpha_{i} \equiv D_{i}/(w_{i}k_{i})$, and $\beta_{i} \equiv a_{i}^{'}/d_{i}^{'}$. At stead state, from $\dot{x_{i}}=0,\dot{y_{i}}=0,(i=1,2)$, we have

\begin{equation}
    \begin{cases}
    \dot{x_{i}}=C_{i}^{\text{(F)}}R^{\text{(F)}}/K_{i},i=1,2; \\
    \dot{y_{i}}=\beta_{i}[C_{i}^{\text{(F)}}]^{2}.
    \end{cases}
\label{xyequ}
\end{equation}
and note that $C_{i} \equiv C_{i}^{\text{(F)}}+x_{i}+2y_{i}$ and $R \equiv R^{\text{(F)}}+\sum\limits_{i=1}^{2}x_{i}$, then,

\begin{subequations}
\begin{align}
& R^{\text{(F)}}=R/(1+C_{1}^{\text{(F)}}/K_{1}+C_{2}^{\text{(F)}}/K_{2}), \label{RCequa} \\
& C_{i}=C_{i}^{\text{(F)}}+R^{\text{(F)}}C_{i}^{\text{(F)}}/K_{i}+2\beta_{i}[C_{i}^{\text{(F)}}]^{2},i=1,2. \label{RCequb}
\end{align}
\end{subequations}
By substituting Eq. \ref{RCequa} into Eq. \ref{RCequb}, we have

\begin{subequations}
\begin{align}
& C_{2}^{\text{(F)}}=\frac{K_{2}}{K_{1}}[\frac{RC_{1}^{\text{(F)}}}{C_{1}-C_{1}^{\text{(F)}}-2\beta_{1}[C_{1}^{\text{(F)}}]^{2}}-K_{1}-C_{1}^{\text{(F)}}] \label{RCa}, \\
& (C_{2}-C_{2}^{\text{(F)}}-2\beta_{2}[C_{2}^{\text{(F)}}]^{2})(1+C_{1}^{\text{(F)}}/K_{1}+C_{2}^{\text{(F)}}/K_{2})=RC_{2}^{\text{(F)}}/K_{2}. \label{RCb}
\end{align}
\end{subequations}
To illustrate the dependencies among the variables, for this moment, we regard $C_{1}$, $C_{2}$ and $R$ as parameters rather than variables. Then, by further substituting Eq. \ref{RCa} into Eq. \ref{RCb}, we get an equation where $C_{1}^{\text{(F)}}$ is the single variable. Thus, we can present $C_{1}^{\text{(F)}}$ with $C_{1}$, $C_{2}$ and $R$, i.e., $C_{1}^{\text{(F)}}=\Phi (C_{1},C_{2},R)$. By further combining with Eqs. \ref{xyequ}, \ref{RCequa} and \ref{RCa}, we can express $C_{2}^{\text{(F)}}, R^{\text{(F)}}, x_{i}$, and $y_{i}$ using $C_{1}$, $C_{2}$ and $R$. Specifically, for $x_{i}$, we have:

\begin{equation}
x_{i}=u_{i}(R,C_{1},C_{2}),i=1,2.
\label{xxequ}
\end{equation}
If all species can coexist, by defining $\Omega_{i}(R,C_{1},C_{2}) \equiv \frac{w_{i}k_{i}}{C_{i}}u_{i}(R,C_{1},C_{2})$ , then, the steady-state equations of $\dot{C_{i}}=0,(i=1,2)$ and $\dot{R}=0$ are:

\begin{equation}
    \begin{cases}
    \Omega_{1}(R,C_{1},C_{2})-D_{1}=0, \\
    \Omega_{2}(R,C_{1},C_{2})-D_{2}=0, \\
    G(R,C_{1},C_{2})=0.
    \end{cases}
\label{omegaG}
\end{equation}
where $G(R,C_{1},C_{2}) \equiv g(R,u_{1}(R,C_{1},C_{2}),u_{2}(R,C_{1},C_{2}),C_{1},C_{2})$. Evidently, Eq. \ref{omegaG} corresponds to three unparallel surfaces and hence share a common point. We verify this conclusion with numerical calculations shown in Figs. 1h and \ref{S8}a-b, where the red dots represent the fixed points. These fixed points can be stable. Therefore, the two consumer species can steadily coexist.

\subsubsection{Analytical solutions of species abundances at steady state}
At steady state, since $\dot{x_{i}}=\dot{y_{i}}=\dot{C_{i}}=0,(i=1,2)$, then,

\begin{equation}
    \begin{cases}
    x_{i}=\alpha_{i}C_{i}, \\
    C_{i}^{\text{(F)}}=K_{i}\alpha_{i}C_{i}/R^{\text{(F)}}, \\
    y_{i}=\beta_{i}(K_{i}\alpha_{i}C_{i})^{2}[R^{\text{(F)}}]^{-2}.\\
    \end{cases}
\label{steadyequ}
\end{equation}
Meanwhile $C_{i}=C_{i}^{\text{(F)}}+x_{i}+2y_{i}$, then, as long as $C_{i},R>0(i=1,2)$, we have

\begin{equation}
C_{i}=\frac{(1-\alpha_{i})[R^{\text{(F)}}]^{2}-K_{i}\alpha_{i}R^{\text{(F)}}}{2\beta_{i}(K_{i}\alpha_{i})^{2}}.
\label{Cequ}
\end{equation}
If the population abundance of resource species is much more than that of the consumers (i.e., $R \gg C_{1}+C_{2}$), then $R \gg x_{1}+x_{2}$ and $R^{\text{(F)}} \approx R$, thus,

\begin{equation}
C_{i}=\frac{(1-\alpha_{i})R^{2}-K_{i}\alpha_{i}R}{2\beta_{i}(K_{i.}\alpha_{i})^{2}}.
\label{Cequapprox}
\end{equation}
We further assume that the population dynamics of the resources follow the same construction rule as that of the MacArthur’s consumer-resource model:

\begin{equation}
    g(R_{l},x_{1},x_{2},C_{1},C_{2})=\begin{cases}
    R_{0}R(1-R/K_{0})-(k_{1}x_{1}+k_{2}x_{2}),\  \text{for\;biotic\;resources} \\
    R_{a}(1-R/K_{0})-(k_{1}x_{1}+k_{2}x_{2}),\  \text{for\;abiotic\;resources} \\
    \end{cases}
\label{ggequa}
\end{equation}
Since $\dot{R}=0$, then, for biotic resources,

\begin{equation}
R=\frac{k_{1}/(2\beta_{1}K_{1})+k_{2}/(2\beta_{2}K_{2})+R_{0}}{\frac{k_{1}(1-\alpha_{1})}{2\beta_{1}\alpha_{1}(K_{1})^{2}}+\frac{k_{2}(1-\alpha_{2})}{2\beta_{2}\alpha_{2}(K_{2})^{2}}+\frac{R_{0}}{K_{0}}}.
\label{Rbio}
\end{equation}
While for abiotic resources,

\begin{equation}
R=\frac{-\kappa_{1}+\sqrt{\kappa_{1}^{2}+4\kappa_{2}R_{a}}}{2\kappa_{2}},
\label{Rabio}
\end{equation}
where $\kappa_{1}=\frac{R_{a}}{K_{0}}-\frac{k_{1}}{2\beta_{1}K_{1}}-\frac{k_{2}}{2\beta_{2}K_{2}}$, and $\kappa_{2}=\frac{k_{1}(1-\alpha_{1})}{2\beta_{1}\alpha_{1}(K_{1})^{2}}+\frac{k_{2}(1-\alpha_{2})}{2\beta_{2}\alpha_{2}(K_{2})^{2}}.$

Eqs. \ref{Cequapprox}, \ref{Rbio}, \ref{Rabio} are the analytical solutions to the steady-state species abundances when $R \gg C_{1}+C_{2}$. As shown in Figure 1e, the analytical solutions agree well with the numerical results (the exact solutions). To conduct a systematic comparison for different model parameters, we assign $D_{i}(i=1,2)$ to be the only parameter of each different value between species $C_{1}$ and $C_{2}$ (with $D_{1}>D_{2}$), and define $\Delta \equiv (D_{1}-D_{2})/D_{2}$ as the competitive difference between the two consumer species. The comparisons between the analytical solutions and the numerical results are shown in Fig. \ref{S8}c-d. Clearly, they are close to each other, exhibiting very good consistency for all cases regardless of the life form of the resources.

Furthermore, we test that if it is achievable to predict the species coexistence region of model parameters with the analytical solutions. As $D_{i}(i=1,2)$ is the only parameter of each different value between the two-consumer species (with $D_{1}>D_{2}$), the supremum (upper limit) of the competitive difference for species coexistence (defined as $\overline{\Delta}$) corresponds to the case when the steady-state solutions of the species abundances 
satisfy $R,C_{2}>0$ and $C_{1}=0^{+}$, where $0^{+}$ stands for the infinitesimal positive number. In the upper surface of the coexistence region, $\Delta=\overline{\Delta}$ and $C_{1}=0^{+}$. Then, using Eq. \ref{Rbio}, and note that $R>0$, we have

\begin{equation}
R=\frac{K_{1}+\alpha_{1}}{1-\alpha_{1}}.
\label{Rbig0}
\end{equation}
Meanwhile,

\begin{equation}
\alpha_{1}=\alpha_{2}(\Delta+1)=\alpha_{2}(\overline{\Delta}+1).
\label{alpha1}
\end{equation}
For biotic resources, combining Eqs. \ref{Rbio}, \ref{Rbig0} and \ref{alpha1}, we have

\begin{equation}
\overline{\Delta}=\frac{1-(K/K_{0}+1)\alpha}{k/(2\beta KR_{0})+(K/K_{0}+1)\alpha},
\label{deltabar}
\end{equation}
where $K=K_{i}$, $k=k_{i}$, and $\beta=\beta_{i}(i=1,2)$. For abiotic resources, combining Eqs. \ref{Rabio}, \ref{Rbig0} and \ref{alpha1}, we have

\begin{equation}
\overline{\Delta}=\frac{1}{\alpha_{2}(K_{1}\rho+1)}-1,
\label{deltab}
\end{equation}
where $\rho=\frac{1}{2}(\frac{1}{K_{0}}-\frac{k_{2}}{2R_{a}\beta_{2}K_{2}})+\frac{1}{2}\sqrt{(\frac{1}{K_{0}}-\frac{k_{2}}{2R_{a}\beta_{2}K_{2}})^{2}+2\frac{k_{2}(1-\alpha_{2})}{R_{a}\beta_{2}\alpha_{2}(K_{2})^{2}}}$

With Eqs. \ref{deltabar} and \ref{deltab}, we can calculate the upper surface of the coexistence region predicted by analytical solutions when $R \gg C_{1}+C_{2}$. The comparisons between analytical predictions and numerical results (exact solution) are shown in Fig. \ref{S8}e-f, which overall exhibits very good consistency.

\subsubsection{Stability analysis and a Hopf bifurcation}

We apply linear stability analysis to study the local stability of the fixed points. Specifically, for a given fixed point $E$ (e.g., $E(x_{1},x_{2},y_{1},y_{2},C_{1},C_{2},R)$), if all the eigenvalues (defined as $\lambda_{i},i=1,\cdots,7$) of the Jacobian matrix $E$ at point are negative 
in the real parts, then, $E$ is locally stable. In contrast, if there exist one or multiple eigenvalues with a nonnegative real part, then the fixed point $E$ is unstable.

In our analysis, when the separation rate $d^{'}(d^{'}=d^{'}_{1}=d^{'}_{2})$ of the interference-pair increases in the vicinity of the critical value $d^{'}_{c}$, the population dynamics of the consumers and resources transits from a stable fixed point into a stable limit cycle (Fig. \ref{S9}e-f, see also Fig. \ref{S9}c-d). In particular, the amplitude of the oscillation gradually increases with the bifurcation parameter $d^{'}$, and quantitatively, the amplitude is proportional to $\sqrt{d^{'}-d^{'}_{c}}$ (Fig. \ref{S9}e). This clearly demonstrates a supercritical Hopf bifurcation.

To further investigate if there exists a non-zero measure parameter space for species coexistence, we set $D_{i}(i=1,2)$ to be the only parameter of each different value between species $C_{1}$ and $C_{2}$, and then $\Delta = (D_{1}-D_{2})/D_{2}$ reflects the completive difference between the two consumer species. As shown in Fig. 2g-h, the region below the blue surface and above the red surface corresponds to stable coexistence, while that below the red surface and above $\Delta=0$ corresponds to unstable fixed points, which may end in a limit cycle or $C_{1}$ extinction. An exemplified transection is shown in Fig. 2i, which corresponds to the $\Delta=1$ plane in Fig. 2h. Above all, there exists a non-zero measure parameter region to promote species coexistence regardless the life form of the resources.

\subsection{$M$ consumers species competing for $N$ resources species}

Here we consider the scenario involving chasing pair and intraspecific interference for the generic case with $M$ types of consumers and $N$ types of resources (see Fig. \ref{S14}a-b). Then, the population dynamics of the consumers and resources can be described as follows:

\begin{equation}
    \begin{cases}
    \dot{x_{il}}=a_{il}C_{i}^{\text{(F)}}R_{l}^{\text{(F)}}-(k_{il}+d_{il})x_{il}, \\
    \dot{y_{i}}=a_{ii}^{'}[C_{i}^{\text{(F)}}]^{2}-d_{ii}^{'}y_{i}, \\
    \dot{C_{i}}=\sum\limits_{l=1}^{N}w_{il}k_{il}x_{il}-D_{i}C_{i}, \\
    \dot{R_{l}}=g_{l}(\{R_{l}\},\{x_{i}\},\{C_{i}\}),i=1,\cdots,M,l=1,\cdots,N. \\
    \end{cases}
\label{multiequ}
\end{equation}
Note that Eq. \ref{multiequ} is identical with Eqs. 1-2, and we use the same variables and parameters as that in the main text. Then, the population of the consumers and resources are $C_{i}=C_{i}^{\text{(F)}}+\sum\limits_{l=1}^{N}x_{il}+2y_{i}$ and $R_{l}=R_{l}^{\text{(F)}}+\sum\limits_{i=1}^{M}x_{il}$. For convenience, we define $K_{il} \equiv (d_{il}+k_{il})/a_{il}, \alpha _{il} \equiv D_{il}/(k_{il}w_{il})$ and $\beta_{i} \equiv a_{ii}^{'}/d_{ii}^{'}(i=1,\cdots,M,l=1,\cdots,N)$.

\subsubsection{Analytical solutions of species abundances at steady state}

At steady state, from $\dot{x_{i}}=0, \dot{y_{i}}=0$, and $\dot{C_{i}}=0$, we have,

\begin{equation}
    \begin{cases}
    x_{il}=C_{i}^{\text{(F)}}R_{l}^{\text{(F)}}/K_{il}, \\
    y_{i}=\beta_{i}[C_{i}^{\text{(F)}}]^{2},\\
    C_{i}^{(F)}=\sum\limits_{l=1}^{N} x_{il}/\alpha_{il}=\sum\limits_{l=1}^{N} C_{i}^{\text{(F)}}R_{l}^{\text{(F)}}/(K_{il}\alpha_{il}). \\
    \end{cases}
\label{multisteadyequ}
\end{equation}
Meanwhile $C_{i}=C_{i}^{\text{(F)}}+\sum\limits_{l=1}^{N}x_{il}+2y_{i}$,  and note that $C_{i}>0$, thus

\begin{equation}
C_{i}^{\text{(F)}}=\frac{1}{2\beta_{i}}[-1+\sum\limits_{l=1}^{N}(\frac{1}{\alpha_{il}}-1)\frac{R_{l}^{\text{(F)}}}{K_{il}}].
\label{CiFequ}
\end{equation}
Combined with Eq. \ref{CiFequ}, and then

\begin{equation}
C_{i}=\sum\limits_{l=1}^{N} \frac{R_{l}^{\text{(F)}}}{2\beta_{i}\alpha_{il}K_{il}}[-1+\sum\limits_{l^{'}=1}^{N}(\frac{1}{\alpha_{il^{'}}}-1)\frac{R_{l^{'}}^{\text{(F)}}}{K_{il^{'}}}].
\label{Ciequ}
\end{equation}
We further assume that the specific function of $g_{l}(\{R_{l}\},\{x_{i}\},\{C_{i}\})$ satisfy Eq. 4, i.e.,

\begin{equation}
    g_{l}(\{R_{l}\},\{x_{i}\},\{C_{i}\})=\begin{cases}
    R_{0}^{(l)}R_{l}(1-R_{l}/K_{0}^{(l)})-\sum\limits_{i=1}^{M}k_{il}x_{il},\  \text{for\;biotic\;resources} \\
    R_{a}^{(l)}(1-R_{l}/K_{0}^{(l)})-\sum\limits_{i=1}^{M}k_{il}x_{il},\  \text{for\;abiotic\;resources} \\
    \end{cases}
\label{gequation}
\end{equation}
For biotic resource, by combining Eqs. \ref{multisteadyequ}, \ref{CiFequ} and \ref{gequation}, we have

\begin{equation}
R_{0}^{(l)}R_{l}(1-\frac{R_{l}}{K_{0}^{(l)}})=\sum\limits_{i=1}^{M} \frac{k_{il}}{2\beta_{i}K_{il}}[-1+\sum\limits_{l^{'}=1}^{N}(\frac{1}{\alpha_{il^{'}}}-1)\frac{R_{l^{'}}^{\text{(F)}}}{K_{il^{'}}}]R_{l}^{\text{(F)}}.
\label{RRequation}
\end{equation}
If the population abundance of each resource species is much more than the total population of all consumers (i.e., $R_{l} \gg \sum\limits_{i=1}^{M} C_{i},l=1,\cdots,N$), then $R_{l} \gg \sum\limits_{i=1}^{M} x_{il}$ and $R_{l}^{\text{(F)}} \approx R_{l}$. Since $R_{l}>0 (l=1,\cdots,N)$, then

\begin{equation}
\sum\limits_{l^{'}=1}^{N}[\delta_{l,l^{'}}\frac{R_{0}^{\text{(l)}}}{K_{0}^{\text{(l)}}}+\sum\limits_{i=1}^{M} \frac{k_{il}}{2\beta_{i}K_{il}K_{il^{'}}}(\frac{1}{\alpha_{il^{'}}}-1)]R_{l^{'}}=R_{0}^{\text{(l)}}+\sum\limits_{i=1}^{M} \frac{k_{il}}{2\beta_{i}K_{il}}.
\label{RRequation2}
\end{equation}
with $\delta_{l,l^{'}}=\begin{cases} 0,\  l \neq l^{'} \\1,\  l=l^{'} \\ \end{cases}$. To present Eq. \ref{RRequation2} in a matrix form, we define matrix $\boldsymbol{A} \equiv [A_{sq}] \in \mathbb{R}^{N \times N}$ ($\mathbb{R}$ stands for the real number field), with

\begin{equation}
A_{sq}=\delta_{s,q}\frac{R_{0}^{\text{(s)}}}{K_{0}^{\text{(s)}}}+\sum\limits_{i=1}^{M} \frac{k_{is}}{2\beta_{i}K_{is}K_{iq}}(\frac{1}{\alpha_{iq}}-1),s,q=1,\cdots,N  
\label{matrixequaa}
\end{equation}
and two arrays

\begin{equation}
\begin{cases} 
\boldsymbol{B} \equiv (R_{0}^{\text{(l)}}+\sum\limits_{i=1}^{M} \frac{k_{i1}}{2\beta_{i}K_{i1}},\cdots, R_{0}^{\text{(N)}}+\sum\limits_{i=1}^{M} \frac{k_{iN}}{2\beta_{i}K_{iN}})^{T}, \\ 
\boldsymbol{R} \equiv (R_{1},\cdots, R_{N})^{T}, \\ 
\end{cases} 
\label{matrixequab}
\end{equation}
where “T” represents the transpose. Then, Eq. \ref{RRequation2} can be written as

\begin{equation}
\boldsymbol{A} \cdot \boldsymbol{R}=\boldsymbol{B} 
\label{matrixequac}.
\end{equation}
We can solve for $\boldsymbol{R}$:

\begin{equation}
\boldsymbol{R}=\frac{adj(\boldsymbol{A})}{det(\boldsymbol{A})} \cdot \boldsymbol{B} 
\label{matrixequad}
\end{equation}
where the $adj(\boldsymbol{A})$ and $det(\boldsymbol{A})$ denote the adjugate matrix and determinant of $\boldsymbol{A}$ respectively. Next, we define $\boldsymbol{C} \equiv (C_{1},\cdots, C_{M})$. Note that $R_{l} \approx R_{l}^{\text{(F)}}$, combined with Eq. \ref{Ciequ}, we have,

\begin{equation}
\begin{cases} 
\boldsymbol{C} \equiv (C_{1},\cdots, C_{M}), \\ 
C_{i}=\sum\limits_{l=1}^{N} \frac{R_{l}}{2\beta_{i}\alpha_{il}K_{il}}[-1+\sum\limits_{l^{'}=1}^{N}(\frac{1}{\alpha_{il^{'}}}-1)\frac{R_{l^{'}}}{K_{il^{'}}}],i=1,\cdots,M. 
\end{cases} 
\label{matrixequae}
\end{equation}
Then, for biotic resources, Eqs. \ref{matrixequaa}-\ref{matrixequae} are the analytical solutions to the steady state species abundances when $R_{l} \gg \sum\limits_{i=1}^{M} C_{i} (l=1,\cdots,N)$. On the other hand, for abiotic resource, by combining Eqs. \ref{multisteadyequ}, \ref{CiFequ} and \ref{gequation}, we have

\begin{equation}
R_{0}^{\text{(l)}}(1-\frac{R_{l}}{K_{0}^{\text{(l)}}})=\sum\limits_{i=1}^{M} \frac{k_{il}}{2\beta_{i}K_{il}}[-1+\sum\limits_{l^{'}=1}^{N}(\frac{1}{\alpha_{il^{'}}}-1)\frac{R_{l^{'}}^{\text{(F)}}}{K_{il^{'}}}]R_{l}^{\text{(F)}}.
\label{Rabioequation}
\end{equation}
If $R_{l} \gg \sum\limits_{i=1}^{M} C_{i} (l=1,\cdots,N)$, then $R_{l} \gg \sum\limits_{i=1}^{M} x_{il}$ and $R_{l}^{\text{(F)}} \approx R_{l}$. Thus,

\begin{equation}
(\frac{R_{0}^{\text{(l)}}}{K_{0}^{\text{(l)}}}-\sum\limits_{i=1}^{M} \frac{k_{il}}{2\beta_{i}K_{il}}+\sum\limits_{l^{'}=1}^{N}\sum\limits_{i=1}^{M} \frac{k_{il}}{2\beta_{i}K_{il}}(\frac{1}{\alpha_{il^{'}}}-1)\frac{R_{l^{'}}^{\text{(F)}}}{K_{il^{'}}})R_{l}=R_{0}^{\text{(l)}},
\label{Rabioequation2}
\end{equation}
with $l=1,\cdots,N$. To present Eq. \ref{Rabioequation2} is a set of second-order ordinary differential equations (ODEs), which is clearly solvable.

Actually, when $N=1, M \geq 1$, and $R_{l} \gg \sum\limits_{i=1}^{M} C_{i} (l=1)$, we can explicitly present the analytical solution to the steady-state species abundances. To simplify the notations, we omit the “l” in the sub-/super-scripts since $N=1$.Then, for biotic resources,

\begin{equation}
\begin{cases} 
R=\frac{R_{0}+\sum\limits_{i=1}^{M} k_{i}/(2\beta_{i}K_{i})}{\sum\limits_{i=1}^{M} \frac{k_{i}(1-\alpha_{i})}{2\beta_{i} \alpha_{i}(K_{i})^{2}}+\frac{R_{0}}{K_{0}}}, \\ 
C_{i}=\frac{1}{2 \beta_{i} \alpha_{i}K_{i}}[(\frac{1}{\alpha_{i}}-1)\frac{R}{K_{i}}-1]R,i=1,\cdots,M. \\ 
\end{cases} 
\label{RCbio}
\end{equation}
For abiotic resources,

\begin{equation}
\begin{cases} 
R=\frac{-\gamma_{1}+\sqrt{\gamma_{1}^{2}+4\gamma_{2}R_{a}}}{2\gamma_{2}}, \\ 
C_{i}=\frac{1}{2\beta_{i} \alpha_{i}K_{i}}[(\frac{1}{\alpha_{i}}-1)\frac{R}{K_{i}}-1]R,i=1,\cdots,M. \\ 
\end{cases} 
\label{RCabio}
\end{equation}
where $\gamma_{1} \equiv \frac{R_{a}}{K_{0}}-\sum\limits_{i=1}^{M} \frac{k_{i}}{2\beta_{i}K_{i}}$ and $\gamma_{2} \equiv \sum\limits_{i=1}^{M} \frac{k_{i}(1-\alpha_{i})}{2\beta_{i} \alpha_{i}(K_{i})^{2}}$. Thus, Eqs. \ref{matrixequaa}-\ref{matrixequae} are the analytical solutions to the steady-state species abundances when $R_{l} \gg \sum\limits_{i=1}^{M} C_{i}$.

Numerically, consistent with the analytical predictions, a handful of resource species ($N$) can support an unexpected wide range of consumers species ($M \gg N$) to coexist at steady state (Figs. 3 and \ref{S14}-\ref{S20}). In the simulations, $D_{i}(i=1,\cdots,M)$is the only parameter of each different value among the consumer species, and thus each consumer species owns a unique competitiveness. The comparisons between the analytical predictions and the ODEs simulation results (exact solution) are shown in Figs. 3a-b and \ref{S14}c-d, which clearly shows a very good consistency. In particular, with stochastic simulation algorithm (SSA), we have further identified that the facilitated biodiversity is resistant to stochasticity (see Figs. 3 and \ref{S15}-\ref{S20}).

\section{Scenario involving chasing pair and interspecific interference}\label{chaintersection}

Here we consider the scenario involving chasing pair and interspecific interference in the case of $M=2$ and $N=1$ (Fig. \ref{S7}a-b), with everything follow that depicted in \ref{sectionchasinter}. 
Then, $C_{i} \equiv C_{i}^{\text{(F)}}+x_{i}+z, R \equiv R^{\text{(F)}}+x_{1}+x_{2}$, and the population dynamics follows (identical with Eq. \ref{interchaseequ}):

\begin{equation}
    \begin{cases}
    \dot{x_{i}}=a_{i}C_{i}^{\text{(F)}}R^{\text{(F)}}-(k_{i}+d_{i})x_{i},i=1,2; \\
    \dot{z}=a_{12}^{'}C_{1}^{\text{(F)}}C_{2}^{\text{(F)}}-d_{12}^{'}z, \\
    \dot{C_{i}}=w_{i}k_{i}x_{i}-D_{i}C_{i}, \\
    \dot{R}=g(R,x_{1},x_{2},C_{1},C_{2}). \\
    \end{cases}
\label{interchaseequ11}
\end{equation}
Here the functional form of $g(R,x_{1},x_{2},C_{1},C_{2})$ is unspecified. For convenience, we define $K_{i} \equiv (d_{i}+k_{i})/a_{i}, \alpha_{i} \equiv D_{i}/(w_{i}k_{i}) (i=1,2)$, and $\gamma \equiv a_{12}^{'}/d_{12}^{'}$. At stead state, from $\dot{x_{i}}=0 (i=1,2)$ and $\dot{z}=0$, we have

\begin{equation}
    \begin{cases}
    \dot{x_{i}}=C_{i}^{\text{(F)}}R^{\text{(F)}}/K_{i},i=1,2; \\
    \dot{z}=\gamma C_{1}^{\text{(F)}}C_{2}^{\text{(F)}}.
    \end{cases}
\label{xzequ11}
\end{equation}
Note that $C_{i} \equiv C_{i}^{\text{(F)}}+x_{i}+z$ and $R \equiv R^{\text{(F)}}+x_{1}+x_{2}$, then,

\begin{equation}
\begin{cases}
    C_{1}=C_{1}^{\text{(F)}}+R^{\text{(F)}}C_{1}^{\text{(F)}}/K_{1}+\gamma C_{1}^{\text{(F)}}C_{2}^{\text{(F)}}, \\
    C_{2}=C_{2}^{\text{(F)}}+R^{\text{(F)}}C_{2}^{\text{(F)}}/K_{2}+\gamma C_{1}^{\text{(F)}}C_{2}^{\text{(F)}}, \\
    R=R^{\text{(F)}}(1+C_{1}^{\text{(F)}}/K_{1}+C_{2}^{\text{(F)}}/K_{2}).
    \end{cases}
\label{C1C2R}
\end{equation}
For this moment, we regard $C_{1}, C_{2}$ and $R$ as parameters rather than variables, then in Eq. \ref{C1C2R}, there are three equations and three variables ($C_{1}^{\text{(F)}}, C_{2}^{\text{(F)}}$ and $R^{\text{(F)}}$) .

Clearly, following similar analysis as that in Sec.\ref{2C1R}, we can present $C_{1}^{\text{(F)}}, C_{2}^{\text{(F)}}$ and $R^{\text{(F)}}$with $C_{1}, C_{2}$ and $R$, e.g., $C_{1}^{\text{(F)}}=\Phi^{'}(C_{1}^{\text{(F)}},C_{2}^{\text{(F)}},R^{\text{(F)}})$. Combined with Eqs. \ref{xzequ11}, we can express $x_{i}$ and $z$using $C_{1}, C_{2}$ and $R$. Specifically, for $x_{i}$, we have

\begin{equation}
x_{i}=u_{i}^{'}(R,C_{1},C_{2}),i=1,2.
\label{xequ11}
\end{equation}
If all species can coexist, by defining $\Omega_{i}^{'}(R,C_{1},C_{2}) \equiv \frac{w_{i}k_{i}}{C_{i}}u_{i}^{'}(R,C_{1},C_{2})$ , then, the steady-state equations of $\dot{C_{i}}=0,(i=1,2)$ and $\dot{R}=0$ are:

\begin{equation}
    \begin{cases}
    \Omega_{1}^{'}(R,C_{1},C_{2})-D_{1}=0, \\
    \Omega_{2}^{'}(R,C_{1},C_{2})-D_{2}=0, \\
    G^{'}(R,C_{1},C_{2})=0.
    \end{cases}
\label{omegaG11}
\end{equation}
where $G^{'}(R,C_{1},C_{2}) \equiv g(R,u_{1}^{'}(R,C_{1},C_{2}),u_{2}^{'}(R,C_{1},C_{2}),C_{1},C_{2})$. Evidently, Eq. \ref{omegaG11} corresponds to three unparallel surfaces and hence share a common point. We verify this conclusion with numerical calculations shown in Figs. 1g and \ref{S6}a-b, where the black dots represent the fixed points. Nevertheless, these fixed points are all unstable, and thus the consumer species still cannot coexist at steady state (Fig. 1d).

\subsection{Analytical results of the fixed-point solutions}

Here we investigate the unstable fixed points where all species coexist ($R,C_{1},C_{2}>0$). 
From ($\dot{x_{i}}=0,(i=1,2), \dot{z}=0, \dot{C_{i}}=0$), and note that $C_{i} \equiv C_{i}^{\text{(F)}}+x_{i}+z$, we have

\begin{equation}
    \begin{cases}
    C_{i}=K_{i}\alpha_{i}C_{i}(R^{\text{(F)}})^{-1}+\alpha_{i}C_{i}+z, i=1,2; \\
    z=\gamma K_{1}\alpha_{1}K_{2}\alpha_{2}(R^{\text{(F)}})^{-2}C_{1}C_{2}. \\
    \end{cases}
\label{steadyequ11}
\end{equation}
Since $C_{i}>0$, then

\begin{equation}
    \begin{cases}
    C_{1}=\frac{(1-\alpha_{2})[R^{\text{(F)}}]^{2}-K_{2}\alpha_{2}R^{\text{(F)}}}{\gamma K_{1}\alpha_{1}K_{2}\alpha_{2}}, \\
    C_{2}=\frac{(1-\alpha_{1})[R^{\text{(F)}}]^{2}-K_{1}\alpha_{1}R^{\text{(F)}}}{\gamma K_{1}\alpha_{1}K_{2}\alpha_{2}}. \\
    \end{cases}
\label{Cequ11}
\end{equation}
If $R \gg C_{1}+C_{2}$, then $R \gg x_{1}+x_{2}$ and $R^{\text{(F)}} \approx R$, we have

\begin{equation}
    \begin{cases}
    C_{1}=\frac{(1-\alpha_{2})R^{2}-K_{2}\alpha_{2}R}{\gamma K_{1}\alpha_{1}K_{2}\alpha_{2}}, \\
    C_{2}=\frac{(1-\alpha_{1})R^{2}-K_{1}\alpha_{1}R}{\gamma K_{1}\alpha_{1}K_{2}\alpha_{2}}. \\
    \end{cases}
\label{Cequ11app}
\end{equation}
We further assume that the population dynamics of the resources follows Eq. \ref{ggequa}. For biotic resources, at the fixed points, $\dot{R}=0$, then

\begin{equation}
R_{0}R(1-\frac{R}{K_{0}})=k_{1}\alpha_{1}C_{1}+k_{2}\alpha_{2}C_{2}.
\label{Rbio11}
\end{equation}
Substituting Eq. \ref{Cequ11app} into Eq. \ref{Rbio11}, and note that $R > 0$, then we have

\begin{equation}
R=\frac{R_{0}+\frac{k_{1}}{\gamma K_{1}}+\frac{k_{2}}{\gamma K_{2}}}{\frac{k_{1}\alpha_{1}+k_{2}\alpha_{2}}{\gamma K_{1}\alpha_{1}K_{2}\alpha_{2}}-\frac{k_{1}+k_{2}}{\gamma K_{1}K_{2}}+\frac{R_{0}}{K_{0}}}.
\label{Rbioequ11}
\end{equation}
For abiotic resources, at the fixed points, $\dot{R}=0$, then

\begin{equation}
R_{a}(1-\frac{R}{K_{0}})=k_{1}\alpha_{1}C_{1}+k_{2}\alpha_{2}C_{2}.
\label{Rabio11}
\end{equation}
Combined with Eq. \ref{Cequ11app}, we have

\begin{equation}
R=\frac{-\kappa_{1}+\sqrt{\kappa_{1}^{2}+4\kappa_{2}R_{a}}}{2\kappa_{2}}.
\label{Rabiosol11}
\end{equation}
where $\kappa_{1}=\frac{R_{a}}{K_{0}}-\frac{k_{1}}{\gamma K_{1}}-\frac{k_{2}}{\gamma K_{2}}$, and $\kappa_{2}=\frac{k_{1}(1-\alpha_{2})}{\gamma K_{1}K_{2}\alpha_{2}}+\frac{k_{2}(1-\alpha_{1})}{\gamma K_{1}K_{2}\alpha_{2}}$

Eqs. \ref{Cequ11app}, \ref{Rbioequ11} and \ref{Rabiosol11} are the analytical results of the fixed-point solutions when $R \gg C_{1}+C_{2}$. As shown in Fig. \ref{S6}c-d, the analytical predictions agree well with the numerical results (exact solutions).

\subsection{Stability analysis of the coexistence behavior}

In the scenario involving chasing pair and interspecific interference, all fixed points are unstable (shown in Fig. \ref{S6}e-f). For abiotic resources, the two-consumer species cannot enduringly coexist (Figs. 1d and \ref{S7}c). For biotic resources, in the deterministic framework, both consumer species may oscillating coexist (Fig. \ref{S7}e, g) or quasi periodic oscillating coexist (Fig. \ref{S7}f, h, see also Fig. \ref{S7} i and j for the Lyapunov exponents and Poincare map). However, within this scenario, for either life form of the resource, the two-consumer species fail to coexist along with stochasticity (Fig. 2a, see also Fig. \ref{S7}l and its ODEs simulation counterpart Fig. \ref{S7}k).

\section{Scenario involving chasing-pairs and both intra- and inter-specific interference}

\subsection{Analytical solutions of species abundances at steady state}

Here we consider the scenario involving chasing-pairs and both intra- and inter-specific interference in the simple case of $M=2$ and $N=1$ (Fig. \ref{S10}a, b, combing that depicted in Sec. \ref{2C1R} and \ref{chaintersection}):

$$\ce{$C_{i}$^{(F)} + $R$^{(F)} <=>[$a_{i}$][$d_{i}$] $C_{i}$^{(P)} \vee $R$^{(P)} ->[$k_{i}$] $C_{i}$^{(F)}(+)}$$
$$\ce{$C_{1}$^{(F)} + $C_{2}$^{(F)} <=>[$a_{12}^{'}$][$d_{12}^{'}$] $C_{1}$^{(P)} \vee $C_{2}$^{(P)}}$$
$$\ce{$C_{i}$^{(F)} + $C_{i}$^{(F)} <=>[$a_{i}^{'}$][$d_{i}^{'}$] $C_{i}$^{(P)} \vee $C_{i}$^{(P)}},i=1,2$$
Then, $C_{i} \equiv C_{i}^{\text{(F)}}+x_{i}+2y_{i}+z$ and $R \equiv R^{\text{(F)}}+x_{1}+x_{2}$, and the population dynamics of the consumers and resources can be described as follows:

\begin{equation}
    \begin{cases}
    \dot{x_{i}}=a_{i}C_{i}^{\text{(F)}}R^{\text{(F)}}-(k_{i}+d_{i})x_{i}, \\
    \dot{z}=a_{12}^{'}C_{1}^{\text{(F)}}C_{2}^{\text{(F)}}-d_{12}^{'}z, \\
    \dot{y_{i}}=a_{i}^{'}[C_{i}^{\text{(F)}}]^{2}-d_{i}^{'}y_{i}, \\
    \dot{C_{i}}=w_{i}k_{i}x_{i}-D_{i}C_{i}, \\
    \dot{R}=g(R,x_{1},x_{2},C_{1},C_{2}),i=1,2. \\
    \end{cases}
\label{intrainterchaseequ}
\end{equation}
where the functional form of $g(R,x_{1},x_{2},C_{1},C_{2})$ follows Eq. \ref{ggequa}. For convenience, we define $K_{i} \equiv (d_{i}+k_{i})/a_{i}, \alpha_{i} \equiv D_{i}/(w_{i}k_{i}), \beta_{i} \equiv a_{i}^{'}/d_{i}^{'}$, and $\gamma \equiv a_{12}^{'}/d_{12}^{'},(i=1,2)$. At stead state, from $\dot{x_{i}}=0,\dot{y_{i}}=0,\dot{z}=0$, and $C_{i}=0,(i=1,2)$, we have

\begin{equation}
    \begin{cases}
    x_{i}=\alpha_{i}C_{i}, \\
    C_{i}^{\text{(F)}}=K_{i}\alpha_{i}C_{i}(R^{\text{(F)}})^{-1}, \\
    y_{i}=\beta_{i}(K_{i}\alpha_{i}C_{i})^{2}[R^{\text{(F)}}]^{-2}, \\
    \dot{z}=\gamma K_{1}\alpha_{1}K_{2}\alpha_{2}[R^{\text{(F)}}]^{-2}C_{1}C_{2}.
    \end{cases}
\label{xyzequ11}
\end{equation}
Combined with $C_{i} \equiv C_{i}^{\text{(F)}}+x_{i}+2y_{i}+z$, and note that $C_{i} > 0 (i=1,2)$, then,

\begin{equation}
    \begin{cases}
    (1-\alpha_{1})(R^{\text{(F)}})^{2}-K_{1}\alpha_{1}R^{\text{(F)}}=2\beta_{1}(K_{1}\alpha_{1})^{2}C_{1}+\gamma K_{1}\alpha_{1}K_{2}\alpha_{2}C_{2}, \\
    (1-\alpha_{2})(R^{\text{(F)}})^{2}-K_{2}\alpha_{2}R^{\text{(F)}}=2\beta_{2}(K_{2}\alpha_{2})^{2}C_{2}+\gamma K_{1}\alpha_{1}K_{2}\alpha_{2}C_{1}.
    \end{cases}
\label{CCRequ}
\end{equation}
If $R \gg C_{1}+C_{2}$, then $R \gg x_{1}+x_{2}$ and we can apply the approximation $R^{\text{(F)}} \approx R$. Combined with Eq. \ref{CCRequ}, and then

\begin{equation}
    \begin{cases}
    C_{1}=R\frac{(2\beta_{2}K_{2}\alpha_{2}(1-\alpha_{1})-\gamma K_{1}\alpha_{1}(1-\alpha_{2}))R+(\gamma-2\beta_{2}) K_{1}\alpha_{1}K_{2}\alpha_{2}}{K_{1}^{2}\alpha_{1}^{2}K_{2}\alpha_{2}(4\beta_{1}\beta_{2}-\gamma^{2})}, \\
    C_{2}=R\frac{(2\beta_{1}K_{1}\alpha_{1}(1-\alpha_{2})-\gamma K_{2}\alpha_{2}(1-\alpha_{1}))R+(\gamma-2\beta_{1}) K_{1}\alpha_{1}K_{2}\alpha_{2}}{K_{1}\alpha_{1}K_{2}^{2}\alpha_{2}^{2}(4\beta_{1}\beta_{2}-\gamma^{2})}.
    \end{cases}
\label{CCRequ2}
\end{equation}
For biotic resources, with $\dot{R}=0$, and note that $R>0$, then we have

\begin{equation}
R=\frac{\frac{k_{1}(\gamma-2\beta_{2})}{K_{1}(4\beta_{1}\beta_{2}-\gamma^{2})}+\frac{k_{2}(\gamma-2\beta_{1})}{K_{2}(4\beta_{1}\beta_{2}-\gamma^{2})}-R_{0}}{\frac{k_{1}2\beta_{2}(\alpha_{1}-1)}{K_{1}^{2}\alpha_{1}(4\beta_{1}\beta_{2}-\gamma^{2})}+\frac{k_{2}2\beta_{1}(\alpha_{2}-1)}{K_{2}^{2}\alpha_{2}(4\beta_{1}\beta_{2}-\gamma^{2})}-\frac{k_{1}\gamma(\alpha_{2}-1)}{K_{1}K_{2}\alpha_{2}(4\beta_{1}\beta_{2}-\gamma^{2})}-\frac{k_{2}\gamma(\alpha_{1}-1)}{K_{1}K_{2}\alpha_{1}(4\beta_{1}\beta_{2}-\gamma^{2})}-\frac{R_{0}}{K_{0}}}.
\label{Rbioequ111}
\end{equation}
While for abiotic resources,

\begin{equation}
R=\frac{-\kappa_{2}^{'}+\sqrt{(\kappa_{2}^{'})^{2}+4\kappa_{1}^{'}R_{a}}}{2\kappa_{1}^{'}},
\label{Rabiosol111}
\end{equation}
where $\kappa_{1}^{'}=\frac{k_{2}\gamma(\alpha_{1}-1)}{K_{1}K_{2}\alpha_{1}(4\beta_{1}\beta_{2}-\gamma^{2})}+\frac{k_{1}\gamma(\alpha_{2}-1)}{K_{1}K_{2}\alpha_{2}(4\beta_{1}\beta_{2}-\gamma^{2})}-\frac{k_{1}2\beta_{2}(\alpha_{1}-1)}{K_{1}^{2}\alpha_{1}(4\beta_{1}\beta_{2}-\gamma^{2})}-\frac{k_{2}2\beta_{1}(\alpha_{2}-1)}{K_{2}^{2}\alpha_{2}(4\beta_{1}\beta_{2}-\gamma^{2})}$, and $\kappa_{2}^{'}=\frac{k_{1}(\gamma-2\beta_{2})}{K_{1}(4\beta_{1}\beta_{2}-\gamma^{2})}+\frac{k_{2}(\gamma-2\beta_{1})}{K_{2}(4\beta_{1}\beta_{2}-\gamma^{2})}+\frac{R_{a}}{K_{0}}$. Eqs. \ref{CCRequ2}-\ref{Rabiosol111} are the analytical solutions to the steady-state species abundances when $R \gg C_{1}+C_{2}$. As shown in Fig. \ref{S11}, the analytical predictions agree well with the numerical results (exact solutions).

\subsection{Stability analysis of the coexisting state}
In the scenario involving chasing pair and both intra- and inter-specific interference, the behavior of species coexistence is very similar to that without interspecific interference. In the deterministic framework, the two-consumer species can coexist either at constant population densities or with time series dynamics such as oscillations (Fig. \ref{S10}f-h). The fixed points can be globally attracting or there is a stable limit cycle (Fig. \ref{S10}c-e and i-k). Clearly, there is a non-zero measure of parameter set where both consumer species can steadily coexist with only one type of resource species (Fig. \ref{S10} i-k). In particular, just as the scenario involving chasing pair and intraspecific interference, the facilitated coexistence state can be maintained along with stochasticity (Fig. \ref{S12}).

Effectively, the influence of interspecific interference is negligible when the separation rate $d_{12}^{'}$ is tremendously large, and vice versa for the intraspecific interference, if the separation rate $d_{i}^{'}(i=1,2)$  is enormous.

\section{Methods}
\subsection{Derivation of the encounter rates with mean-field approximations}
In the model scenario depicted in Fig. 1a, freely consumer and resource individuals move randomly in space, and we can regard the movements as Brownian motions. Specifically, at moment $t$, a consumer individual of species $C_{i} (i=1,\cdots,M)$ moves at speed $v_{C_{i}}$ and with velocity $\boldsymbol{v}_{C_{i}}(t)$, while a resource individual of species $R_{l} (l=1,\cdots,N)$ moves at speed and with velocity $v_{R_{l}}$ and with velocity $\boldsymbol{v}_{R_{l}}(t)$. Here $v_{C_{i}}$ and $v_{R_{l}}$ are two time invariants, while the directions of $\boldsymbol{v}_{C_{i}}(t)$ and $\boldsymbol{v}_{R_{l}}(t)$ change constantly. At moment $t$, the relative velocity is $\boldsymbol{u}_{C_{i}-R_{l}}(t) \equiv \boldsymbol{v}_{R_{l}}(t) - \boldsymbol{v}_{C_{i}}(t)$. We denote the relative speed as $u_{C_{i}-R_{l}}(t)$ and use $\theta(t)$ to represent the angle between $\boldsymbol{v}_{C_{i}}(t)$ and $\boldsymbol{v}_{R_{l}}(t)$. Evidently, $(u_{C_{i}-R_{l}}(t))^{2}=v_{C_{i}}^{2}+v_{R_{l}}^{2}-2v_{C_{i}}v_{R_{l}} \cdot cos\theta(t)$. Since the system is homogenous, then, $\overline{cos \theta}=0$ (the overline means time average), and the average relative speed is $\overline{u_{C_{i}-R_{l}}}=\sqrt{v_{C_{i}}^{2}+v_{R_{l}}^{2}}$. Similarly, the average relative speed between two consumer individuals (of species $C_{i}$ and $C_{j}$, respectively, with $i,j=1,\cdots,M)$) is $\overline{u_{C_{i}-C_{j}}}=\sqrt{v_{C_{i}}^{2}+v_{C_{j}}^{2}}$. Clearly, $\overline{u_{C_{i}-C_{i}}}=\sqrt{2}v_{C_{i}}$.

Next, we apply the mean-field approximations to calculate encounter rates $a_{il}$ (among individuals of species $C_{i}$ and $R_{l}$) and $a_{il}^{'}$ (among individuals of species $C_{i}$ and $C_{j}$), which in essence is the same method in statistical physics applied to calculate the mean free path of gas particles. For convenience, we denote the concentrations of consumer species $C_{i}$ and resource species $R_{l}$ as $n_{C_{i}}$ and $n_{R_{l}}$. Then, $n_{C_{i}}=C_{i}/L^{2}$ and $n_{R_{l}}=R_{l}/L^{2}$. Likewise, we can obtain the concentration of the freely wandering part of both species: $n_{C_{i}^{\text{(F)}}}=C_{i}^{\text{(F)}}/L^{2}$ and $n_{R_{l}^{\text{(F)}}}=R_{l}^{\text{(F)}}/L^{2}$.

In the well-mixed system, consider that all individuals of resource species $R_{l}$ stand still, while a consumer individual (of species $C_{i}$) moves randomly at speed $u_{C_{i}-R_{l}}(t)$ (Fig. \ref{S1}). For a given time interval $\Delta t$ (corresponds to a macroscopic short, while microscopic long interval in statistical physics), the number of encounters between the given consumer individual and freely individuals from resource species $R_{l}$ can be approximated by $2r_{il}^{(C)}n_{R^{\text{(F)}}}\overline{u_{C_{i}-R}}\Delta t$ ($r_{il}^{(C)}$ represents the radius to form a chasing pair, see Fig.1a). Then, for all freely individuals of species $C_{i}$, the total number of encounters with $R^{(F)}$ in interval $\Delta t$ is $\frac{2r_{il}^{(C)}\overline{u_{C_{i}-R}}C_{i}^{\text{(F)}}R^{\text{(F)}}}{L^{2}}\Delta t$. Meanwhile, in the ODEs representation, this corresponds to $a_{i}C_{i}^{\text{(F)}}R^{\text{(F)}}\Delta t$. Comparing both terms above, evidently, for chasing pair, we have $a_{il}=2r_{il}^{(C)}L^{-2}\overline{u_{C_{i}-R_{l}}}=2r_{il}^{(C)}L^{-2}\sqrt{v_{C_{i}}^{2}+v_{R_{l}}^{2}}$. Likewise, 
for interspecific interference, we have $a_{il}^{'}=2r_{il}^{(I)}L^{-2}\overline{u_{C_{i}-C_{j}}}=2r_{il}^{(I)}L^{-2}\sqrt{v_{C_{i}}^{2}+v_{C_{j}}^{2}}$, while 
for intraspecific interference, we have $a_{ii}^{'}=2\sqrt{2}v_{C_{i}}r_{ii}^{(I)}L^{-2}$

\subsection{Stochastic simulations and Individual-based modeling}

To consider the impact of stochasticity on species coexistence, we apply stochastic simulation algorithm (SSA) \cite{gillespie2007stochastic} and individual-based modeling (IBM) \cite{grimm2013individual, vetsigian2017diverse} to simulate the stochastic process. For the SSA, we follow the Gillespie’s standard algorithm and the simulation procedures.

For the IBM, we consider a 2D system of squared landscape in a length of $L$ with periodic boundary conditions, and only for the case of $M = 2$ and $N = 1$. Consumer $C_{i} (i=1,2)$ individuals move at speed $v_{C_{i}}$, while resource $R$ individuals move at speed $v_{R}$. In our simulations, the unit length is $\Delta l=1$, and all the populations move probabilistically. For instance, when $\Delta t$ is very small ($v_{C_{i}}\Delta t \ll 1$), a $C_{i}$ individual moves a unit length with probability $v_{C_{i}}\Delta t$. Specifically, we simulate the time evolution of the model system following the procedures below.

\emph{Initialization.} The initial point of each individual is chosen randomly from a uniform distribution in the squared landscape. For convenience, we only consider the points with both integers in the $x$ and $y$ coordinates.

\emph{Moving.} The destination of a movement is chosen randomly among four directions ($x$ -positive, $x$ -negative, $y$ -positive, $y$ -negative) following a uniform distribution. Then consumer $C_{i}$ individuals move $\Delta l$ with probability $v_{C_{i}}\Delta t$, while resource individuals move $\Delta l$ with probability $v_{R}\Delta t$.

\emph{Forming pairs.} When a consumer $C_{i}$ individual and a resource individual get close in space within a distance of $r_{i}^{(C)}$, the two individuals form a chasing-pair. Likewise, when two consumer individuals $C_{i}$ and $C_{j}$ stand within a distance of $r_{ij}^{(I)}$, they form a chasing-pair. 

\emph{Dissociating pairs.} In the simulations, we update the system with small time step $\Delta t$ so that $d_{i}\Delta t,k_{i}\Delta t \ll 1$. Then, a random number $\widetilde{\varsigma}$ is chosen from a uniform distribution between 0 and 1. If $\widetilde{\varsigma}$ is smaller than the $d_{i}\Delta t(i=1,2)$, then, the pair dissociates into the two separated individuals. One individual occupies the same position as the previous pair, while the other individual gets just out of the encounter radius in a random angle that is uniformly distributed. In the consumption process, if $\widetilde{\varsigma}$ is greater than $d_{i}\Delta t$ yet smaller than $(d_{i}+k_{i})\Delta t(i=1,2)$, then the biomass of the resource flows into the consumer populations (updated according to the birth procedure), while the consumer individual occupies the same position as the previous pair and then updated following the moving procedure. Finally, if $\widetilde{\varsigma}$ is greater than $(d_{i}+k_{i})\Delta t(i=1,2)$, the pair maintain its current status.

\emph{Birth and death.} In each time step of the updates, the birth and death of each species accumulates, and we count them using a positive number with decimals. The integer part of this number will be updated in this run if it is no less than 1. A newborn is updated following the initialization procedure. The death process is also chosen randomly from the living species.

The simulation parameters of Figure 1-4 are as follows:

In 1c, f: $a_{i} = 0.1, d_{i} = 0.5, w_{i} = 0.1 , k_{i} = 0.1 , i =1, 2, D_{1}=0.002, D_{2}=0.001, K_{0}=5, R_{a} = 0.05$.
In 1d, g: $a_{i} = 0.05 , a_{ij}^{'} = 0.3 , d_{i} = 0.5 , d_{ij}^{'} = 0.01 , w_{i} = 0.08 , k_{i} = 0.02 , i, j =1, 2, i \neq j, D_{1}=0.001, D_{2}=0.0009, K_{0}=10, R_{a} = 0.1$. 
In 1e, h: $a_{i} = 0.5 , a_{i}^{'} = 0.525 , d_{i} = 0.5 , d_{i}^{'} = 0.5 , w_{i} = 0.2 , k_{i} = 0.4 , i =1, 2 , D_{1}=0.022, D_{2}=0.020, K_{0}=10, R_{0} = 0.1$.

In 2a-b: $K_{0} = 60, R_{0} = 0.05, a_{12}^{'}=0.02, d_{12}^{'}=0.02$. In (i): $K_{0} = 60, R_{0} = 0.05$. 
In 2c: $a_{i}  = 0.15, d_{i}  = 0.1, k_{i} = 0.2, w_{i} = 0.1, D_{1} = 0.0009, D_{2} = 0.0007, K_{0} = 60, R_{0} = 0.15$.  
In 2d: $a_{i} = 0.1 , a_{i}^{'} = 0.125 , d_{i} = 0.1 , d_{i}^{'} = 0.05 , w_{i} = 0.1 , k_{i} = 0.1 , i =1, 2 , D_{1}=0.0035, D_{2}=0.0038, K_{0}=100, R_{a} = 0.3$.
In 2e-i: $a_{i} = 0.1 , d_{i} = 0.1 ,  w_{i} = 0.1 , k_{i} = 0.1 , i =1, 2 , K_{0}=100 $. In 2e: $ a_{i}^{'} = 0.125 , d_{i}^{'} = 0.05, D_{1}=0.0085, D_{2}=0.0080, R_{0} = 0.05$. In 2f: $ a_{i}^{'} = 0.125 , d_{i}^{'} = 0.1D_{1}=0.0085, D_{2}=0.0080, R_{0} = 0.05$
In 2g: $ D_{2}=0.001, \Delta=(D_{1}-D_{2})/D_{2}, K_{0}=100, R_{a} = 0.1$.
In 2h: $ D_{2}=0.001, \Delta=(D_{1}-D_{2})/D_{2}, K_{0}=100, R_{0} = 0.05$.
In 2i: $ D_{2}=0.001, \Delta=1, K_{0}=100, R_{0} = 0.05$.
In 2j, m:  $a_{i} = 0.02 , a_{i}^{'} = 0.025 , d_{i} = 0.7 , d_{i}^{'} = 0.7 , w_{i} = 0.4 , k_{i} = 0.05 , i =1, 2 , D_{1}=0.0160, D_{2}=0.0171, K_{0}=2000, R_{a} = 5.5$.  
In 2k-l:  $a_{i} = 0.06 , a_{i}^{'} = 0.075 , d_{i} = 2 , d_{i}^{'} = 2 , w_{i} = 0.32 , k_{i} = 0.22 , i =1, 2 , D_{1}=0.0550, D_{2}=0.0551, K_{0}=5000, R_{0} = 0.13$.  
In 2m-o: $ L = 120, r = 5, v_{C} = v_{R} = 1, l_{1} = 50 , l_{2} = 50 , a_{i} = 0.0039l_{1} , a_{i}^{'} = 0.0039l_{2} , d_{i}^{'} = 0.4,  d_{i} = 0.4 , w_{i}= 0.3 , k_{i} = 0.1, i =1, 2, D_{1} = 0.0080, D_{2} = 0.0085, K_{0} = 200, R_{0} = 0.5$.

In 3a: $ a_{il} = 0.05, a_{il}^{'} = 0.07, d_{il} = 1.05 , d_{il}^{'} = 0.018,w_{il} = 0.45 , k_{il} = 0.16 , i =1,\cdots,5, l=1,2,3, K_{0}^{(1)}=600, K_{0}^{(2)}=1000, K_{0}^{(3)}=800,  R_{0}^{(1)}=R_{0}^{(2)}=0.9, R_{0}^{(3)}=0.95, D_{1}=0.062, D_{2}=0.0615, D_{3}=0.0639, D_{4}=0.066, D_{5}=0.0644$.  
In 3b-c: $a_{il} = 0.1 , a_{i}^{'} = 0.125 , d_{il} = 0.5 , d_{i}^{'} = 0.3, w_{il}= 0.2 , k_{il} = 0.2, R_{0}^{(l)}=0.95, R_{0}^{(2)}=0.85, R_{0}^{(3)}=0.9, K_{0}^{(1)}=6000, K_{0}^{(2)}=4000, K_{0}^{(3)}=5000, D_{i}=0.03+0.005 \times  \xi_{i}$ ($\xi_{i}$ is a random number between 0 and 1), $i=1,\cdots,18, l=1,2,3$.  
In 3d: $a_{i} = 0.1 , a_{i}^{'} = 0.125 , d_{i} = 0.3 , d_{i}^{'} = 0.05 , w_{i} = 0.35 , k_{i} = 0.5 , i =1,\cdots,5 , D_{1}=0.0320, D_{2}=0.0335, D_{3}=0.0345, D_{4}=0.0350, D_{5}=0.0360, K_{0}=3000, R_{a} = 0.35$. 
In 3e-f: $a_{i} = 0.1 , d_{i} = 0.5 , w_{i}= 0.1 , k_{i} = 0.1 , D_{i}=0.001+0.001 \times \xi_{i}$ ($\xi_{i}$ is a random number between 0 and 1), $i=1,\cdots,20, K_{0} =10000, R_{a} = 5, a_{i}^{'} = 0.125 , d_{i}^{'} = 0.1$.  
In 3g-i: $ a_{i} = 0.1 , a_{l}^{'} = 0.125 , d_{i} = 0.5 , d_{i}^{'} = 0.2, w_{i}= 0.2 , k_{i} = 0.1 , K_{0}=10^{5}, D_{i}= $ Normal$(1, 0.38)\times 0.008,  i=1,\cdots,200, R_{a}=150. $  

In 4a-b:  $a_{i} = 0.1 , a_{i}^{'} = 0.125, d_{i} = 0.105,  w_{i} = 0.2 , k_{i} = 0.1 , i =1, 2, D_{1}=0.0110, K_{0}=2000, R_{a} = 15.5. $
In 4a: $ d_{i}^{'} = 0.1525, D_{2}=0.0143$.  
In 4b: $ d_{i}^{'} = 0.2, D_{2}=0.0154$. 
In 4c-d: $a_{i} = 0.1 , a_{i}^{'} = 0.125,  d_{i}^{'} = 0.2, k_{i} = 0.1 , i =1, 2, K_{0}=1000, R_{a} = 2. $
In 4c: $w_{i} = 0.2, d_{i} = 0.5, D_{1} = 0.0060, D_{2} = 0.0075$.  
In 4d: $w_{i} = 0.1, d_{i} = 0.8, D_{1} = 0.0024, D_{2} = 0.0028$.

\clearpage
\section*{Supplemental Figures}
\begin{figure}[ht!]   
\centering
\includegraphics[width=0.5\textwidth]{FigS1.pdf}
\caption{\label{S1} Estimation of the pairwise encounter rate with mean-field approximations. To calculate the average collisions frequency, supposing that all the populations stand still except for one individual (e.g., a consumer from species $C_{i}$). Over a short time interval $\Delta t$, this consumer moves from the middle left to the upper right of the depicted region (a very small part of the whole system) following the center of the two parallel dashed lines. Meanwhile it encounters many individuals from, e.g., resource species $R_{l}$. Then we can estimate the collisions frequency and thus the encounter rate $a_{il}$ by counting the number of $R_{l}$ individuals within the area between the two dashed lines. $\rho$ stands for the radius of encounter.}   
\end{figure}

\begin{figure}[ht!]   
	\centering
	\includegraphics[width=0.8\textwidth]{FigS2.pdf}
	\caption{\label{S2} Functional response in the scenario involving only chasing pair. (a-b) The red surface corresponds to the B-D model (calculated with Eq.\ref{FXiCP4}), while the green surface represents the exact solutions to our mechanistic model (using Eq. \ref{FXiCP1}), and the magenta (using Eq. \ref{FXiCP2}) and blue (using Eq. \ref{FXiCP3}) surfaces represent the approximate solutions to our model. In (a-b): $k=0.1, a = 0.25$. In (a): $d = 0$. In (c): $k = 0.5, a = 0.025$.}   
\end{figure}

\begin{figure}[ht!]   
	\centering
	\includegraphics[width=0.8\textwidth]{FigS3.pdf}
	\caption{\label{S3} Functional response in the scenario involving chasing pair and intraspecific interference. (a-b) The red surface corresponds to the B-D model (calculated with Eq.\ref{FXiABD}), while the green surface represents the exact solutions to our mechanistic model (using Eq. \ref{FXiA1}), and the blue surface (with Eq. \ref{FXiA2}) and the magenta surface (with Eq. \ref{FXiA3}) represent the quasi-rigorous and the approximate solutions to our model, respectively. In (a-b): $a = 0.1 , k = 0.1 , d^{'}=0.1, a^{'}=0.12.$}   
\end{figure}

\begin{figure}[ht!]   
	\centering
	\includegraphics[width=0.8\textwidth]{FigS4.pdf}
	\caption{\label{S4} Functional response in the scenario involving chasing pair and interspecific interference. (a-b) The red surface corresponds to the B-D model (calculated with Eq.\ref{F12ABD}), while the green surface represents the quasi-rigorous solutions to our mechanistic model (using Eq. \ref{FXi1}), and the blue surface (using Eq. \ref{FXi2}) represents the approximate solutions to our model. In (a-c): $a_{1}=a_{2} = 0.1 , k_{1}=k_{2} = 0.1 , d_{12}^{'}=0.1, a_{12}^{'}=0.6$.}   
\end{figure}

\begin{figure}[ht!]   
\centering
\includegraphics[width=0.8\textwidth]{FigS5.pdf}
\caption{\label{S5} Chasing-pair scenario is under the constraint of competitive exclusion. (a) If all consumer species coexist at steady state, $f_{i}(R^{\text{(F)}})/D_{i}=1 (i=1, 2)$, where $f_{i}(R^{\text{(F)}})\equiv R^{\text{(F)}}/(R^{\text{(F)}}+K_{i})=D_{i}$, with $K_{i}=(d_{i}+k_{i})/a_{i}$. This means that the three lines $f_{i}(R^{\text{(F)}})/D_{i}=1, i=1, 2$ and $y=1$ share a common point, which is generally impossible except for special parameter settings. (b) The blue plane is parallel to the green one, and hence they do not have a common point. (c-d) Time courses of the species abundances in the scenario involving only chasing pair. The two consumer species cannot enduringly coexist. In (c): $a_{1} = a_{2} = 0.1, k_{1} = k_{2} = 0.1, w_{1} = w_{2} = 0.1, d_{1} = d_{2} = 0.5, D_{1} = 0.002, D_{2} = 0.001, K_{0} = 5, R_{a} = 0.05$. In (d): $a_{1} = a_{2} = 0.1, k_{1} = k_{2} = 0.05, w_{1} = w_{2} = 0.1, d_{1} = d_{2} = 0.2, D_{1} = 0.005, D_{2} = 0.004, R_{0} = 0.05, K_{0} = 100$.}   
\end{figure}

\begin{figure}[ht!]   
	\centering
	\includegraphics[width=0.75\textwidth]{FigS6.pdf}
	\caption{\label{S6} Fixed point solutions in the case of $M = 2$ and $N = 1$ involving chasing pair and interspecific interference. $D_{i} (i=1,2)$ is the only parameter of each different value between the consumer species, and $\Delta=(D_{1}-D_{2})/D_{2}$ represents the competitive differences between them. (a-b) Positive solutions to the steady-state equations: $\dot{R_{1}}=0$ (orange surface), $\dot{C_{1}}=0$ (blue surface), $\dot{C_{2}}=0$ (green surface). The intersection point marked by black dots are unstable fixed points. (a-b) were calculated using Eqs. S4.1, S3.10. (c-d) Comparisons between numerical results and analytical solutions of the steady-state species abundances in this system. Color bars are analytical solutions while hollow bars are numerical results. The numerical results (labeled with superscript ‘Numerical’) were calculated from Eqs. S3.1 and S3.10, while the analytical solutions (labeled with superscript ‘Analytical’) were calculated from Eqs. S4.10, S4.12. (e-f) In this scenario, there is no parameter space for steady coexistence. The region below the red surface and above $\Delta=0$ represents unstable fixed points, which may end in a limit cycle (see Fig. \ref{S7}e, g), a torus (see Fig. \ref{S7}f, h), or $C_{1}$ extinction (see Fig. \ref{S7}c-d, $D_{1}>D_{2}$). In (a): $a_{1} = a_{2} = 0.05, d_{1} = d_{2} = 0.05, K_{0} = 20, a_{12}^{'}=0.3, d_{12}^{'}=0.01, k_{1} = k_{2} = 0.02, w_{1} = w_{2} = 0.08, D_{1} = 0.001, D_{2} = 0.0009, R_{a} = 0.01$. In (b): $a_{1} = a_{2} = 0.05, d_{1} = d_{2} = 0.05, K_{0} = 5, a_{12}^{'}=0.3, d_{12}^{'}=0.1, k_{1} = k_{2} = 0.02, w_{1} = w_{2} = 0.08, D_{1} = 0.001, D_{2} = 0.0008, R_{0} = 0.02$. In (c): $a_{1} = a_{2} = 0.04, d_{1} = d_{2} = 0.2, K_{0} = 10, a_{12}^{'}=0.6, d_{12}^{'}=0.1, k_{1} = k_{2} = 0.1, w_{1} = w_{2} = 0.3, D_{2} = 0.0008, R_{a} = 0.2$. In (d): $a_{1} = a_{2} = 0.05, d_{1} = d_{2} = 0.2, K_{0} = 10, a_{12}^{'}=0.6, d_{12}^{'}=0.02, k_{1} = k_{2} = 0.1, w_{1} = w_{2} = 0.2, D_{2} = 0.005, R_{0} = 0.02$. In (e-f): $a_{1} = a_{2} = 0.05, d_{1} = d_{2} = 0.1, K_{0} = 100, k_{1} = k_{2} = 0.1, w_{1} = w_{2} = 0.05, D_{2} = 0.0005$. In(e): $R_{a}=0.5$; In(f): $R_{0}=0.05$.}   
\end{figure}

\begin{figure}[ht!]   
	\centering
	\includegraphics[width=0.75\textwidth]{FigS7.pdf}
	\caption{\label{S7} Interspecific interference may facilitate oscillating coexistence without stochasticity. (a-b) Model scenario involving chasing-pairs and interspecific interference in the case of $M=2$ and $N=1$. (c-d) The simulate results of species abundances for abiotic or biotic resources. The two-consumer species cannot coexist at steady state. (e-f) In the ODEs simulations, both consumer species may coexist with time series dynamics. However, in the SSA simulations where we apply the same set of parameters, the two-consumer species do not coexist. (g-h) In a 3D phase space, the ODEs simulation in (e-f) corresponds to a limit cycle or a quasi-periodic torus. (i-j) The Lyapunov exponent analysis in (i) and Poincare map in (j) further suggest that the dynamics in (f) and (h) is a quasi-periodic oscillation. In (i), $L_{1}, L_{2}, \cdots, L_{6}$ represents the Lyapunov exponents. Then, in the Lyapunov spectrum, three exponents are zeros while the rest are all negative, which clearly suggest a 3-D torus. In (j) The loops in the Poincare map indicate a quasi-periodic oscillation. (k) In the ODEs simulations, the cyan region represents oscillating coexistence while the magenta region represents species extinction. (l) In the SSA simulation results, which share the parameter region as that in (k), there is no area for species coexistence. In (c-j): $a_{1} = a_{2} = 0.05, d_{1} = d_{2} = 0.1, k_{1} = k_{2} = 0.1, w_{1} = w_{2} = 0.05, D_{1} = 0.0009, D_{2} = 0.0007$. In (c): $K_{0} = 10, R_{a} = 0.2, a_{12}^{'}=0.02, d_{12}^{'}=0.02$. In (d): $K_{0} = 10, R_{0} = 0.05, a_{12}^{'}=0.02, d_{12}^{'}=0.02$. In (e, g): $K_{0} = 60, R_{0} = 0.05, a_{12}^{'}=0.06, d_{12}^{'}=0.02$. In (f, h, e, j): $K_{0} = 60, R_{0} = 0.05, a_{12}^{'}=0.02, d_{12}^{'}=0.02$. In (i): $K_{0} = 60, R_{0} = 0.05$. In (k-l): $a_{1} = a_{2} = 0.15, d_{1} = d_{2} = 0.1, k_{1} = k_{2} = 0.2, w_{1} = w_{2} = 0.1, D_{1} = 0.0009, D_{2} = 0.0007, K_{0} = 60, R_{0} = 0.15$.}   
\end{figure}

\begin{figure}[ht!]   
	\centering
	\includegraphics[width=0.75\textwidth]{FigS8.pdf}
	\caption{\label{S8} Fixed point solutions in the case of $M = 2$ and $N = 1$ involving intraspecific interference. $D_{i} (i=1,2)$ is the only parameter of each different value between the consumer species, and $\Delta=(D_{1}-D_{2})/D_{2}$ represents the competitive differences between them. (a-b) Positive solutions to the steady-state equations: $\dot{R_{1}} = 0$ (orange surface), $\dot{C_{1}} = 0$ (blue surface), $\dot{C_{2}} = 0$ (green surface). The intersection point marked by red dots are stable fixed points. (a-b) were calculated with Eqs. \ref{intrachaseequ2} and \ref{ggequa}. (c-d) Comparisons between numerical results and analytical solutions of the steady-state species abundances in this system. Color bars are analytical solutions while hollow bars are numerical results. The numerical results (labeled with superscript ‘Numerical’) were calculated from Eqs. \ref{intrachaseequ2} and \ref{ggequa}, 
		while the analytical solutions (labeled with superscript ‘Analytical’) were calculated from Eqs. \ref{Cequapprox}, \ref{Rbio}and \ref{Rabio}. (e-f) Comparison between the numerical results and analytical solutions of the coexistence region. Here represents the maximum tolerated for species coexistence. The red and cyan surfaces represent the analytical solutions (calculated with Eqs. \ref{deltabar}, \ref{deltab}) and numerical results (calculated with Eqs. \ref{intrachaseequ2}, \ref{ggequa}), respectively. In (a-b) $a_{1} = a_{2} = 0.5, a_{1}^{'}=a_{2}^{'} = 0.625 , d_{1} = d_{2} = 0.5, d_{1}^{'} = d_{2}^{'} = 0.5, k_{1} = k_{2} = 0.4, D_{2} = 0.02, w_{1} = w_{2} = 0.5, K_{0} = 10$. In (a): $D_{1} = 1.2D_{2}, R_{a}= 0.1$. In (b): $D_{1} = 1.05D_{2}, R_{0} = 0.3$. In (c): $a_{1} = a_{2} = 0.1, a_{1}^{'}=a_{2}^{'} = 0.12, k_{1} = k_{2} = 0.12, w_{1} = w_{2} = 0.3, D_{2} = 0.02, K_{0} = 100, R_{a} = 0.8, d_{1} = d_{2} = 0.5, d_{1}^{'} = d_{2}^{'} = 0.05$. In (d): $a_{1} = a_{2} = 0.05, a_{1}^{'}=a_{2}^{'} = 0.06, k_{1} = k_{2} = 0.12, w_{1} = w_{2} = 0.2, D_{2}= 0.008, K_{0} = 100, R_{0} = 0.1, d_{1} = d_{2} = 0.8, d_{1}^{'} = d_{2}^{'} = 0.01$. In (e): $a1 = a2 = 0.5, k_{1} = k_{2} = 0.2, w_{1} = w_{2} = 0.2, D_{2} =0.008, K_{0} = 60, R_{a} = 0.8, d_{1} = d_{2} = 0.8$. In (f): $a_{1} = a_{2} = 0.5, k_{1} = k_{2} = 0.1, w_{1} = w_{2} = 0.2, D_{2} = 0.008, K_{0} = 100, R_{0} = 0.2, d_{1} = d_{2} = 0.8$.}   
\end{figure}

\begin{figure}[ht!]   
	\centering
	\includegraphics[width=0.8\textwidth]{FigS9.pdf}
	\caption{\label{S9} Numerical results in the scenario involving chasing pair and intraspecific interference. (a-b) Model scenario in the case of $M=2$ and $N=1$. (c-d) The time series dynamics exhibit two types of coexisting behavior. The two-consumer species may coexist either at steady state or with oscillating behavior. (e) The phase diagram of Hopf bifurcation when varying parameter $d^{'}(d^{'} \equiv d_{1}^{'}=d_{2}^{'})$. (f) The Hopf bifurcation 
		curve when varying parameter $d^{'}$. (g) The two-consumer species can enduringly coexist along with stochasticity (h) The stochastic coexistence state is stable and globally attractive (see (g) for the time courses). In (c-f): $a_{1}=a_{2} = 0.1 , a_{1}^{'}=a_{2}^{'} = 0.125 , d_{1}=d_{2} = 0.1 , k_{1}=k_{2} = 0.1, w_{1}=w_{2} = 0.1, K_{0} = 100, D_{1} = 0.0085, D_{2} = 0.008, R_{0} = 0.05.$ In (c): $d_{1}^{'}=d_{2}^{'}=0.05$. In (d): $d_{1}^{'}=d_{2}^{'}=0.1$. In (g-h): $a_{1}=a_{2} = 0.06 , a_{1}^{'}=a_{2}^{'} = 0.075 , d_{1}=d_{2} = 2 , d_{1}^{'}=d_{2}^{'}=2,  k_{1}=k_{2} = 0.22, w_{1}=w_{2} = 0.32, K_{0} = 500, D_{1} = 0.055, D_{2} = 0.057, R_{0} = 0.13.$}   
\end{figure}

\begin{figure}[ht!]   
	\centering
	\includegraphics[width=0.75\textwidth]{FigS10.pdf}
	\caption{\label{S10} Numerical results in the scenario involving chasing pair and both intra- and inter-specific interference. The species coexistence behavior is similar to that without interspecific interference. (a-b) Model scenario in the case of $M = 2$ and $N = 1$ (c-e) 3D Phase diagram of species coexistence region. $D_{i} (i=1,2)$ is the only parameter of each different value between the consumer species, and then $\Delta=(D_{1}-D_{2})/D_{2}$ measures the competitive difference between the two species. The parameter region below the blue surface yet above the red surface represents species stable coexistence, while that below the red surface and above $\Delta=0$ represents unstable fixed points. (e) The transection corresponding to the plane $\Delta=0.2$ in (d). The blue, cyan and magenta region represent stable coexistence, oscillating coexistence and $C_{1}$ extinction, respectively. (f-h) Time series of the species coexistence either at constant population densities (f-g) or with oscillations. (i-j) The coexistence state is globally stable attractors. (k) The coexistence state is globally unstable, and all trajectories attract to a stable limit cycle. In (c): $a_{1} = a_{2} = 0.1, a_{1}^{'} = a_{2}^{'} = 0.12, k_{1} = k_{2} = 0.1, w_{1} = w_{2} = 0.1, D_{2} = 0.001, d_{1} = d_{2} = 0.3, K_{0} = 100, a_{12}^{'}=0.05, R_{a} = 0.3$. In (f, i):  $a_{1} = a_{2} = 0.1, a_{1}^{'} = a_{2}^{'} = 0.12, k_{1} = k_{2} = 0.2, w_{1} = w_{2} = 0.1, D_{1} = 0.0009, D_{2} = 0.0085, d_{1} = d_{2} = 0.2, d_{1}^{'} = d_{2}^{'} = 0.3, K_{0} = 100, a_{12}^{'}=0.05, d_{12}^{'}=0.2, R_{a} = 0.9$. In (d-e, g-h i-k): $a_{1} = a_{2} = 0.1, a_{1}^{'} = a_{2}^{'} = 0.14, k_{1} = k_{2} = 0.2, w_{1} = w_{2} = 0.05, D_{1} = 0.0009, D_{2} = 0.0085, d_{1} = d_{2} = 0.2, d_{1}^{'} = d_{2}^{'} = 0.3, K_{0} = 100, a_{12}^{'}=0.05, R_{0} = 0.1$. In (e): $\Delta=0.2$. In (g, j): $d_{12}^{'}=0.2$; In(h, k): $d_{12}^{'}=0.4$.}   
\end{figure}

\begin{figure}[ht!]   
	\centering
	\includegraphics[width=0.85\textwidth]{FigS11.pdf}
	\caption{\label{S11}  Comparisons between numerical results and analytical solutions of the steady-state species abundances in the scenario involving chasing pair and both intra- and inter-specific interference. $D_{i} (i=1,2)$ is the only parameter of each different value between the consumer species, and $\Delta=(D_{1}-D_{2})/D_{2}$ measures the competitive differences between the two species. Color bars are analytical solutions while hollow bars are numerical results. The numerical results (labeled with superscript ‘Numerical’) were calculated from Eqs. S5.1, S3.10, while the analytical solutions (labeled with superscript ‘Analytical’) were calculated from Eqs. S5.4-S5.6. In (a): $a_{1} = a_{2} = 0.05, a_{1}^{'} = a_{2}^{'} = 0.06, k_{1} = k_{2} = 0.1, w_{1} = w_{2} = 0.2, D_{2} = 0.008, K_{0} = 100, R_{a} = 0.8, d_{1} = d_{2} = 0.5, d_{1}^{'} = d_{2}^{'} = 0.002, a_{12}^{'}=0.2, d_{12}^{'}=0.2$. In (b):  $a_{1} = a_{2} = 0.05, a_{1}^{'} = a_{2}^{'} = 0.06, k_{1} = k_{2} = 0.05, w_{1} = w_{2} = 0.2, D_{2} = 0.006, K_{0} = 100, R_{0} = 0.2, d_{1} = d_{2} = 0.5, d_{1}^{'} = d_{2}^{'} = 0.002, a_{12}^{'}=0.2, d_{12}^{'}=0.2$.} 
\end{figure}

\begin{figure}[ht!]   
	\centering
	\includegraphics[width=0.85\textwidth]{FigS12.pdf}
	\caption{\label{S12} Outcomes of two types consumers species competing for single resource species involving chasing pair and intra- and inter-specific interference. (a-b) Both consumer species can coexist with either lifeform of the resources regardless of stochasticity. In (a): $a_{1} = a_{2} = 0.1, a_{1}^{'}=a_{2}^{'} = 0.11 , k_{1} = k_{2} = 0.1, w_{1} = w_{2} = 0.15, D_{1} = 0.0125, D_{2} = 0.012, K_{0} = 300, R_{a} = 0.8, d_{1} = d_{2} = 0.3, d_{1}^{'} = d_{2}^{'} = 0.5, a_{12}^{'}=0.01, d_{12}^{'}=0.8$. In (b): $a_{1} = a_{2} = 0.1, a_{1}^{'}=a_{2}^{'} = 0.11 , k_{1} = k_{2} = 0.1, w_{1} = w_{2} = 0.2, D_{1} = 0.013, D_{2} = 0.0125, K_{0} = 500, R_{0} = 0.2, d_{1} = d_{2} = 0.3, d_{1}^{'} = d_{2}^{'} = 0.5, a_{12}^{'}=0.05, d_{12}^{'}=0.4$.}   
\end{figure}

\begin{figure}[ht!]   
\centering
\includegraphics[width=0.7\textwidth]{FigS13.pdf}
\caption{\label{S13} Stochasticity jeopardizes species coexistence. (a-b) We simulate Koch’s model \cite{koch1974competitive} and Huisman-Weissing model \cite{huisman1999biodiversity} using stochastic simulation algorithm (SSA) with identical parameters as their deterministic model. Nevertheless, the two deterministic cases of oscillating coexistence fail as stochasticity is introduced. See Ref. \cite{koch1974competitive} and \cite{huisman1999biodiversity} for the simulation parameters.}   
\end{figure}

\begin{figure}[ht!]   
\centering
\includegraphics[width=0.7\textwidth]{FigS14.pdf}
\caption{\label{S14} With intraspecific interference, a single biotic resource species ($N = 1$) can support 5 consumers species ($M = 5$) to coexist at steady state. (a-b) is a simplified version of Fig. 1(a-b), where the interspecific interference is omitted. (c-d) Time courses of the species abundances simulated with ODEs. Here, $D_{i} (i=1,\cdots,M)$ is the only parameter of each different value among the consumer species, and thus each consumer species owns a unique competitiveness. The dotted lines in (c-d) are the analytical solutions at steady state (calculated from Eqs. \ref{RRequation2}-\ref{RCabio}). In (c): $a_{i} = 0.04 , a_{i}^{'} = 0.056 , d_{i} = 0.6 , d_{i}^{'} = 0.04 , w_{i} = 0.45 , k_{i} = 0.15 , i =1,\cdots,5 , D_{1}=0.0619, D_{2}=0.0595, D_{3}=0.057, D_{4}=0.0584, D_{5}=0.0603, K_{0}=400, R_{a} = 0.9$. In (d): $a_{il} = 0.05 , a_{i}^{'} = 0.07 , d_{il} = 1.05 , d_{i}^{'} = 0.018,w_{il} = 0.45 , k_{il} = 0.16 , i =1,\cdots,5, l=1,2,3, K_{0}^{(1)}=600, K_{0}^{(2)}=1000, K_{0}^{(3)}=800,  R_{0}^{(1)}=R_{0}^{(2)}=0.9, R_{0}^{(3)}=0.95, D_{1}=0.062, D_{2}=0.0615, D_{3}=0.0639, D_{4}=0.066, D_{5}=0.0644$.}   
\end{figure}

\begin{figure}[ht!]   
\centering
\includegraphics[width=0.85\textwidth]{FigS15.pdf}
\caption{\label{S15} With intraspecific interference, one type of abiotic resource species ($N= 1$) can support a wide range of consumers species ($M = 20$) to enduringly coexist regardless of stochasticity. Here, $D_{i} (i=1,\cdots,M)$ is the only parameter of each different value among the consumer species, so each consumer species owns a unique competitiveness. (a-b) Time series of the consumer and resource species simulated with ODEs. (c-d) Time series of the consumer and resource species simulated with SSA (with the same parameters as that in (a-b)). (a, c) With only chasing-pair, consumer species cannot coexist. (b, d) With chasing-pair and intraspecific interference, all consumer species can enduringly coexist regardless of stochasticity. In (a-d): $a_{i} = 0.1 , d_{i} = 0.5 , w_{i}= 0.1 , k_{i} = 0.1 , D_{i}=0.001+0.001 \times \xi_{i}$ ($\xi_{i}$ is a random number between 0 and 1), $i=1,\cdots,20,  K_{0} =10000, R_{a} = 5$. In (a, c): $a_{i}^{'} = 0 , d_{i}^{'} = 0$. In (b, d): $a_{i}^{'} = 0.125 , d_{i}^{'} = 0.1$}   
\end{figure}

\begin{figure}[ht!]   
\centering
\includegraphics[width=0.85\textwidth]{FigS16.pdf}
\caption{\label{S16} With intraspecific interference, one type of biotic resource species ($N= 1$) can support a wide range of consumers species ($M = 20$) to enduringly coexist regardless of stochasticity. Here, $D_{i} (i=1,\cdots,M)$ is the only parameter of each different value among the consumer species, so each consumer species owns a unique competitiveness. (a-b) Time series of the consumer and resource species simulated with ODEs. (c-d) Time series of the consumer and resource species simulated with SSA (same parameters as that in (a-b)). (a, c) With only chasing-pair, consumer species cannot coexist. (b, d) With chasing-pair and intraspecific interference, all consumer species can coexist regardless of stochasticity. In (a-d): $a_{i} = 0.1 , d_{i} = 0.3 , R_{0} = 0.95, w_{i}= 0.1 , k_{i} = 0.1 , D_{i}=0.004+0.002 \times \xi_{i}$ ($\xi_{i}$ is a random number between 0 and 1), $i=1,\cdots,20, K_{0} =300$. In (a, c): $a_{i}^{'} = 0 , d_{i}^{'} = 0$. In (b, d): $a_{i}^{'} = 0.125 , d_{i}^{'} = 0.3$.}   
\end{figure}

\begin{figure}[ht!]   
\centering
\includegraphics[width=0.85\textwidth]{FigS17.pdf}
\caption{\label{S17} In the scenario involving chasing-pair and intraspecific interference, one type of resource species ($N= 1$) can support an unexpected wide range of consumers species ($M = 100$) to enduringly coexist regardless of stochasticity. Here, $D_{i} (i=1,\cdots,M)$ is the only parameter of each different value among the consumer species, so each consumer species owns a unique competitiveness. (a-b) Time series of ODEs simulations. (c-d) Time series of SSA simulations, which share the same parameter set as that in (a-b). In (a, c): $a_{i} = 0.1 , a_{i}^{'} = 0.125 , d_{i} = 0.3 , d_{i}^{'} = 0.3, w_{i}= 0.3 , k_{i} = 0.1 , R_{a} = 50, K_{0} =10000, D_{i}=0.002+0.002 \times \xi_{i}$ ($\xi_{i}$ is a random number between 0 and 1), $i=1,\cdots,100$. In (b, d): $a_{i} = 0.1 , a_{i}^{'} = 0.125 , d_{i} = 0.5 , d_{i}^{'} = 0.1, w_{i}= 0.1 , k_{i} = 0.1 , R_{0} = 0.95, K_{0} =1000, D_{i}=0.002+0.005 \times \xi_{i}$ ($\xi_{i}$ is a random number between 0 and 1), $i=1,\cdots,100$.}   
\end{figure}

\begin{figure}[ht!]   
\centering
\includegraphics[width=0.85\textwidth]{FigS18.pdf}
\caption{\label{S18} In the scenario involving chasing-pair and intraspecific interference, a handful of resource species ($N = 3$) can support a wide range of consumers species ($M = 18$) to enduringly coexist regardless of stochasticity. Here, $D_{i} (i=1,\cdots,M)$ is the only parameter of each different value among the consumer species, so each consumer species owns a unique competitiveness. (a-b) Time series of ODEs simulations. (c-d) Time series of SSA simulations, which share the same parameter set as that in (a-b). In (a, c): $a_{il} = 0.1 , a_{i}^{'} = 0.125 , d_{il} = 0.5 , d_{i}^{'} = 0.1, w_{il}= 0.2 , k_{il} = 0.2 , K_{0}^{(1)}=8000, K_{0}^{(2)}=3000, K_{0}^{(3)}=5000, D_{i}=0.028+0.008 \times  \xi_{i}$ ($\xi_{i}$ is a random number between 0 and 1), $i=1,\cdots,18, l=1,2,3, R_{a}^{(1)}=30, R_{a}^{(2)}=40, R_{a}^{(3)}=25$. In (b, d): $a_{il} = 0.1 , a_{i}^{'} = 0.125 , d_{il} = 0.5 , d_{i}^{'} = 0.3, w_{il}= 0.2 , k_{il} = 0.2 , R_{0}^{(l)}=0.95, R_{0}^{(2)}=0.85, R_{0}^{(3)}=0.9, K_{0}^{(1)}=6000, K_{0}^{(2)}=4000, K_{0}^{(3)}=5000, D_{i}=0.03+0.005 \times  \xi_{i}$ ($\xi_{i}$ is a random number between 0 and 1), $i=1,\cdots,18, l=1,2,3$.}   
\end{figure}

\begin{figure}[ht!]   
\centering
\includegraphics[width=0.85\textwidth]{FigS19.pdf}
\caption{\label{S19} In the scenario involving chasing-pair and intraspecific interference, a handful of resource species ($N= 3$) can support an unexpected wide range of consumers species ($M = 98$) to enduringly coexist regardless of stochasticity. Here, $D_{i} (i=1,\cdots,M)$ is the only parameter of each different value among the consumer species, so each consumer species owns a unique competitiveness. (a-b) Time series of ODEs simulations. (c-d) Time series of SSA simulations, which share the same parameter set as that in (a-b). In (a, c): $a_{il} = 0.1 , a_{l}^{'} = 0.125 , d_{il} = 0.5 , d_{l}^{'} = 0.3, w_{il}= 0.3 , k_{il} = 0.2 , K_{0}^{(1)}=8000, K_{0}^{(2)}=3000, K_{0}^{(3)}=5000, D_{i}=0.01+0.005 \times \xi_{i}$ ($\xi_{i}$ is a random number between 0 and 1), $i=1,\cdots,98, l=1,2,3, R_{a}^{(1)}=30, R_{a}^{(2)}=40, R_{a}^{(3)}=25$. In (b, d): $a_{il} = 0.2 , a_{i}^{'} = 0.25 , d_{il} = 0.4 , d_{i}^{'} = 0.2, w_{il}= 0.3 , k_{il} = 0.3 , R_{0}^{(l)}=0.85, R_{0}^{(2)}=0.95, R_{0}^{(3)}=0.9, K_{01}=1800, K_{02}=1400, K_{03}=1600, D_{i}=0.008+0.01 \times \xi_{i}$ ($\xi_{i}$ is a random number between 0 and 1), $i=1,\cdots,98, l=1,2,3$.}   
\end{figure}

\begin{figure}[ht!]  
\centering
\includegraphics[width=0.95\textwidth]{FigS20.pdf}
\caption{\label{S20} In the scenario involving chasing-pair and intraspecific interference, a single of resource species ($N = 1$) can support an unexpected wide range of consumers species ($M = 200$, $500$) to enduringly coexist regardless of stochasticity. Here, $D_{i} (i=1,\cdots,M)$ is the only parameter of each different value among the consumer species, so each consumer species owns a unique competitiveness. (a-b) Time series of ODEs simulations. (d-e) Time series of SSA simulations, which share the same parameter set as that in (a-b). (c, f) Rank-species abundance of consumer species. Black dots are experiment data see Ref. \cite{Ubiquitous2018} (see experiment data TARA\_139.SUR.180.2000.DNA and TARA\_054.SUR.180.2000.DNA). Red dots are obtained by numerically Eqs. (1-2, 4) up to time $t = 10^{5}$ at stable state. Blue dots are SSA results corresponding to red dots. Here, we assume that the death rate $D_{i} (i=1,\cdots,M)$ obey a normal distribution (Normal($ \mu $, $ \sigma $)) with mean $ \mu $ and standard deviation $ \sigma $. (g-h) Survival consumers species at different times corresponding to (b, e, case seed1). (i) Rank-species abundance of consumer species at different times corresponding to (b). In (a-c, g, i): $a_{i} = 0.1 , a_{l}^{'} = 0.125 , d_{i} = 0.5 , d_{i}^{'} = 0.3, w_{i}= 0.2 , k_{i} = 0.1 , K_{0}=10^{6}, D_{i}=$ Normal$(1, 0.37)\times 0.006,  i=1,\cdots,200, R_{a}=200,$. In (d-f, h): $a_{i} = 0.1 , a_{l}^{'} = 0.125 , d_{i} = 0.3 , d_{i}^{'} = 0.3, w_{i}= 0.2 , k_{i} = 0.1 , K_{0}=2\times10^{6}, D_{i}=$ Normal$(1, 0.35)\times 0.005,  i=1,\cdots,500, R_{a}=200$.}   
\end{figure}

\begin{figure}[ht!]   
\centering
\includegraphics[width=0.75\textwidth]{FigS21.pdf}
\caption{\label{S21} Time series which corresponds to Fig. 4 in the long-term behavior. The simulation details are the same as Fig. 4.}
\end{figure}

\begin{figure}[ht!]   
\centering
\includegraphics[width=0.75\textwidth]{FigS22.pdf}
\caption{\label{S22} Coexistence region in the case of $M = 2$ and $N = 1$ involving intraspecific interference. $D_{i} (i = 1, 2)$ is the only parameter of each different value between the consumer species, and $\Delta=(D_{1}-D_{2})/D_{2}$ represents the competitive differences between them. In (a-b): $ a_{i} = 0.1 , a_{i}^{'} = 0.125 , d_{i} = 0.5 , w_{i}= 0.1 , k_{i} = 0.1 , K_{0} = 100, R_{a} = 5, D_{2} = 0.0014, i = 1, 2. $ }   
\end{figure}

\begin{figure}[ht!]   
\centering
\includegraphics[width=0.85\textwidth]{FigS23.pdf}
\caption{\label{S23} A model of intraspecific interference explains two classical experimental studies that invalidating the CEP. Here, the solid dots, triangles and boxes represent the time serials experimental data, which are connected by the dotted lines for the sake of visibility. The dashed lines and solid lines stand for the ODEs and IBM simulation results, respectively. In all cases, two consumer species ($M = 2$) enduringly coexist with only one type of resources ($N = 1$). (a-b) In Ayala’s experiment [45], two \textit{Drosophila} species (consumers), \textit{D. serrata} and \textit{D. pseudoobscura}, compete for the same type of abiotic resources in a laboratory bottle. (c-d) In Park’s experiment [46], two Tribolium species, \textit{T. confusum} and \textit{T. castaneum}, compete for the same food (flour). In (a-b): $ L = 120, r = 5, v_{C} = v_{R} = 1, l_{1} = 50 , l_{2} = 10 , a_{i} = 0.0039l_{1} , a_{i}^{'} = 0.0039l_{2} , d_{i}^{'} = 0.02, D_{2} = 0.007, K_{0} = 200, R_{a} = 0.5,$ In (a): $d_{i} = 0.01 , w_{i}= 0.3 , k_{i} = 0.15 , D_{1} = 0.01;$ In (b): $d_{i} = 0.02 , w_{i}= 0.29 , D_{1} = 0.014, i = 1, 2.$ In (c-d): $ L = 100, r = 5, v_{C} = v_{R} = 1, l_{1} = 50 , l_{2} = 100 , a_{i} = 0.0057l_{1} , a_{i}^{'} = 0.0057l_{2} ,  w_{i}= 0.4 , d_{i}^{'} = 0.01, D_{1} = 0.007, D_{2} = 0.006, d_{i} = 0.3911 , R_{a} = 2.$ In (c): $k_{i} = 0.13 , K_{0} = 150;$ In (d): $k_{i} = 0.12 , K_{0} = 160, i = 1, 2. $ }   
\end{figure}

\begin{figure}[ht!]   
\centering
\includegraphics[width=0.85\textwidth]{FigS24.pdf}
\caption{\label{S24} Time series which corresponds to Fig. S23 in the long-term behavior. The simulation details are the same as Fig. S23.}   
\end{figure}

\begin{figure}[ht!]   
\centering
\includegraphics[width=0.95\textwidth]{FigS25.pdf}
\caption{\label{S25} Intraspecific interference results in a negative feedback and thus promotes biodiversity. (a-b) As the population density of consumer species $C_{i}$ increases, a larger fraction of species $C_{i}$ are involved in an intraspecific interference pair which temporarily absent from hunting. Meanwhile, the fraction of species $C_{i}$  within a chasing pair decrease. (c) The formation of intraspecific interference virtually leads to a negative feedback and thus promotes biodiversity. In (a): $a_{i} = 0.2 , a_{i}^{'} = 0.25 , d_{i} = 0.3 , d_{i}^{'} = 0.05, w_{i}= 0.5 , k_{i} = 0.15 , R_{a} = 0.35, K_{0} =3000, D_{i}=0.03+i\times 5\times 10^{-4}, i=1,\cdots,20$. In (b): $a_{i} = 0.008 , a_{i}^{'} = 0.0112 , d_{i} = 0.6 , d_{i}^{'} = 0.04, w_{i}= 0.45 , k_{i} = 0.15 , R_{0} = 0.8, K_{0} =400, D_{i}=0.047+i\times 2.5\times 10^{-4}, i=1,\cdots,20$.}   
\end{figure}

\bibliographystyle{plain}
\bibliography{references}